\documentclass[aps,prd,onecolumn,nofootinbib,superscriptaddress, 12pt]{revtex4}
\usepackage[pdftex]{hyperref}
\usepackage{graphicx}% Include figure files
\usepackage{amsmath,slashed}

\usepackage{epstopdf}
\usepackage{subfigure}
\usepackage{verbatim}
\def\ltap{\ \raise.3ex\hbox{$<$\kern-.75em\lower1ex\hbox{$\sim$}}\ }
\def\gtap{\ \raise.3ex\hbox{$>$\kern-.75em\lower1ex\hbox{$\sim$}}\ }
\newcommand{\gsim}{\lower.7ex\hbox{$\;\stackrel{\textstyle>}{\sim}\;$}}
\newcommand{\lsim}{\lower.7ex\hbox{$\;\stackrel{\textstyle<}{\sim}\;$}}
\def\OO{{\cal O}}
\def\LL{{\cal L}}

\newcommand{\GeV}{\,\mathrm{GeV}}

\newcommand{\keV}{\,\mathrm{keV}}

\def\unit{\relax{\rm 1\kern-.26em I}}

\newcommand{\esc}{{\text{esc}}}

\newcommand{\kpc}{{\text{ kpc}}}
\newcommand{\kms}{{\text{ km/s}}}
\newcommand{\vmin}{v_{\text{min}}}

\begin{document}

% Page numbers bottom-center
\pagestyle{plain}

\title{
\begin{flushright}
\mbox{\normalsize FERMILAB-PUB-12-108-A}\\
%\mbox{\normalsize SLAC-PUB-XXXXX}
\end{flushright}
\vskip 15 pt
Dark Matter in 3D}

\author{Daniele S. M. Alves}
\affiliation{Fermi National Accelerator Laboratory, Batavia,
IL 60510}

\author{ Sonia El Hedri}
\affiliation{SLAC National Accelerator Laboratory, Menlo Park, CA 94025}
\affiliation{Stanford Institute for Theoretical Physics, Stanford University, Stanford, CA 94305 }

\author{Jay G. Wacker}
\affiliation{SLAC National Accelerator Laboratory, Menlo Park, CA 94025}
\affiliation{Stanford Institute for Theoretical Physics, Stanford University, Stanford, CA 94305 }

\date{\today}

\begin{abstract}
We discuss the relevance of directional detection experiments in the post-discovery era and propose a method to extract the local dark matter phase space distribution from directional data. The first feature of this method is a parameterization of the dark matter distribution function in terms of integrals of motion, which can be analytically extended to infer properties of the global distribution if certain equilibrium conditions hold. The second feature of our method is a decomposition of the distribution function in moments of a model independent basis, with minimal reliance on the ansatz for its functional form. We illustrate our method using the Via Lactea II N-body simulation as well as an analytical model for the dark matter halo. We conclude that $\OO(1000)$ events are necessary to measure deviations from the Standard Halo Model and constrain or measure the presence of anisotropies.

\end{abstract}
\pacs{} \maketitle
\pagebreak
\tableofcontents

\section{Introduction}

Dark matter comprises 80\% of the mass of the Universe and its presence has been inferred gravitationally in numerous different physical settings including galactic rotation curves,
%CHANGED removed references in this paragraph \cite{Rubin1970}
precision studies of the cosmic microwave background radiation, and through gravitational lensing of galaxies and clusters of galaxies.
%CHANGED removed references in this paragraph \cite{Bacon2003}.  
 However, dark matter's non-gravitational interactions remain unknown and there is an active search underway to determine dark matter's identity.    
  
 Many well motivated proposals to extend the Standard Model accommodate naturally a dark matter candidate that is stable, electrically neutral and has the correct relic abundance \cite{Jungman:1995df}. These models generically predict that dark matter should interact with Standard Model particles with electroweak-sized cross sections. Currently, a large experimental effort is underway searching for interactions of  dark matter with nuclei in underground experiments.

Typically, direct detection experiments attempt to infer that a dark matter particle scattered off a nucleus by detecting the nuclear recoil and measuring the energy deposited in the detector.  
%Given that many ordinary processes can cause nuclear recoils in the same energy range where dark matter is expected, these experiments are performed underground with radio-purified material, in order to reduce cosmic ray fluxes and avoid natural sources of radiation.
%
%Among the generic features that make an experiment competitive are excellent control of backgrounds, sensitivity to low nuclear recoil energies and large exposures (obtained by a combination of large detector mass and long exposure times).
%
%Other features such as the properties of the nuclear target (such as mass, spin, etc) also affect the experimental sensitivity to different classes of dark matter particles. Unfortunately, those are model dependent and one cannot know {\it a priori} what particle physics determines dark matter interactions with the standard model. Therefore it is of fundamental importance to have variety among direct detection experiments, both in the choice of nuclear target as well as in the experimental techniques used to search for nuclear recoils.
%
While most presently running experiments were designed to only measure the energy of the nuclear recoil, more can be learned about dark matter if the direction of nuclear recoils is measured as well \cite{Spergel:1987kx}.  The anisotropy of dark matter-induced nuclear recoils partially arises from the motion of the Earth through galactic dark matter halo, but can also arise from intrinsic anisotropies in the dark matter halo \cite{Kuhlen:2009vh}. Therefore, measuring the direction of the nuclear recoils is not only a powerful handle to discriminate signal from background but also provides a whole class of new measurements of the dark matter's velocity distribution in the solar neighborhood \cite{Copi:1999pw,Green:2006cb,Sciolla:2008vp,Billard:2009mf,Mayet:2010dq,Billard:2011kt}.
%CHANGED from "Therefore, measuring the direction of the nuclear recoil is both a powerful handle to discriminate signal from background and also provides a whole class of new measurements about the velocity distribution of the dark matter halo \cite{Copi:1999pw,Green:2006cb,Sciolla:2008vp,Billard:2009mf,Mayet:2010dq,Billard:2011kt}." 

Over the past decade, the first experiments capable of detecting the directionality of nuclear recoils have been built and commissioned \cite{Ahlen:2009ev,Monroe:2011er,Santos:2011kf,Morgan:2003qp,Martoff:1996cd,SnowdenIfft:1999hz,Shimizu:2002ik,Munoz:2003gx,Sekiya:2004fw,Vahsen:2011qx}.  More recently, the first prototypes have started to release results  that are approaching 0.1 kg-day exposures \cite{Ahlen:2010ub,Miuchi:2007jy,Miuchi:2007ga,Miuchi:2010hn}.  

The majority of the upcoming directional detection experiments use time projection chambers (TPCs) to resolve the direction of the nuclear recoils, by means of drifting the ionized track originated from the nuclear recoils through an electric field and projecting it over a charge collection grid. Two of the dimensions are then inferred from this projection, while the third dimension is obtained by the drift time of the particles in the ionized track. In order to resolve the direction of the track to good accuracy, these TPCs have to operate at low pressure and density, and therefore are limited in target mass, as well as in volume, since the drift length is limited by diffusion \cite{Morgan:2003qp,Shimizu:2002ik}.
The limitation in exposure of directional detection experiments makes them  currently  not competitive with current experiments such as XENON100 \cite{Aprile:2011hi} and CDMS \cite{Ahmed:2009zw}, amongst many others. 

%In order to compensate for that, these experiments aim for large sensitivity to spin-dependent interactions, by choosing nuclear targets with large spin, such as $\text{CF}_4$ and $\text{CS}_2$.

Most likely, directional detection experiments will not be able to scale their exposures sufficiently fast to be the first to discover dark matter. However, they will provide unique measurements of dark matter properties, once dark matter is discovered  through direct detection. The unique capabilities of directional detection experiments have been under-emphasized. For instance, directional detection experiments will explore the kinematics of the interaction between dark matter and Standard Model nuclei more directly than direct detection experiments \cite{Finkbeiner:2009ug,Mayet:2011nk,Cerdeno:2010jj,Green:2003ud,Green:2010ri,Vasquez:2012px}. 

This article will explore how directional detection experiments can measure astrophysical properties of dark matter, such as its phase space distribution in the solar neighborhood. Previous studies in this domain have been carried out in \cite{Green:2010gw,Peter:2011wi,Peter:2009ak,Drees:2007hr,Lee:2012pf} assuming specific parameterizations for the functional form of the velocity distribution. One important goal of this article is to investigate the dependence of the dark matter phase space distribution in different coordinates, such as integrals of motion. If the dark matter distribution in the solar neighborhood is dominated by the equilibrium component, measuring the dependence of the local distribution on integrals of motion will allow us to infer properties of the global distribution. Another goal of this article is to remain agnostic about the functional form of the local distribution and explore generic parameterizations in terms of a basis of special functions.

More properties of the entire galactic halo can be inferred by combining the measurement of the phase space distribution function with N-body simulations, ultimately  giving clues to the original formation of the Milky Way galaxy. 
 
The organization of this article is as follows.  In Sec.~\ref{Sec: Rate} we overview general aspects of directional detection in the context of post-discovery. In Sec.~\ref{Sec: Jeans} we discuss Jeans theorem and how it motivates integrals of motion as powerful variables in terms of which the dark matter distribution function can be expressed.  In Sec.~\ref{decomposition} we present a novel method to parameterize the functional form of the distribution function - instead of relying on ansatzes of its functional form, we perform a special function decomposition of the dark matter phase space distribution. The coefficients of this decomposition will be the measurable quantities in a directional experiment. We perform fits of these expansion coefficients in Sec.~\ref{Sec: VDF} by simulating data from a mock experiment, and discuss the level of precision in the measurement the velocity distribution function that can be achieved with plausible numbers of events at upcoming directional detection experiments. 
In Sec.~\ref{Sec: VLII} we test our method using the Via Lactea II N-body simulation as a template of the dark-matter distribution in the solar neighborhood, by scattering the particles of the simulation at a detector placed at 8kpc from the center. We then use this mock signal to recover the input distribution, providing a closure/bias test for the procedure.
Finally, in Sec.~\ref{Sec: Discussion} we discuss the implications for upcoming experiments.

\section{General Aspects of Directionality}
\label{Sec: Rate}

In this section we discuss how the direction of nuclear recoils captures the underlying dynamics of the interaction between dark matter and nuclei. We will also show how anisotropies in the dark matter velocity distribution are manifest in rate anisotropies.

\subsection{Directional dependence from form factors and kinematics}

This subsection reviews the basic kinematics of dark matter direct detection. Consider a dark matter particle with mass $m_\text{dm}$ and initial velocity $\vec{v}_\text{dm}$ scattering off of a nucleus at rest with mass $m_N$. The angle between the nuclear recoil direction and $\vec{v}_\text{dm}$, $\eta$, is given by:
\begin{equation}
\label{eqkine}
\cos\eta=\frac{v_\text{min}(E_R)}{v_\text{dm}}.
\end{equation}
\begin{figure}[t]
\includegraphics[width=6.2in]{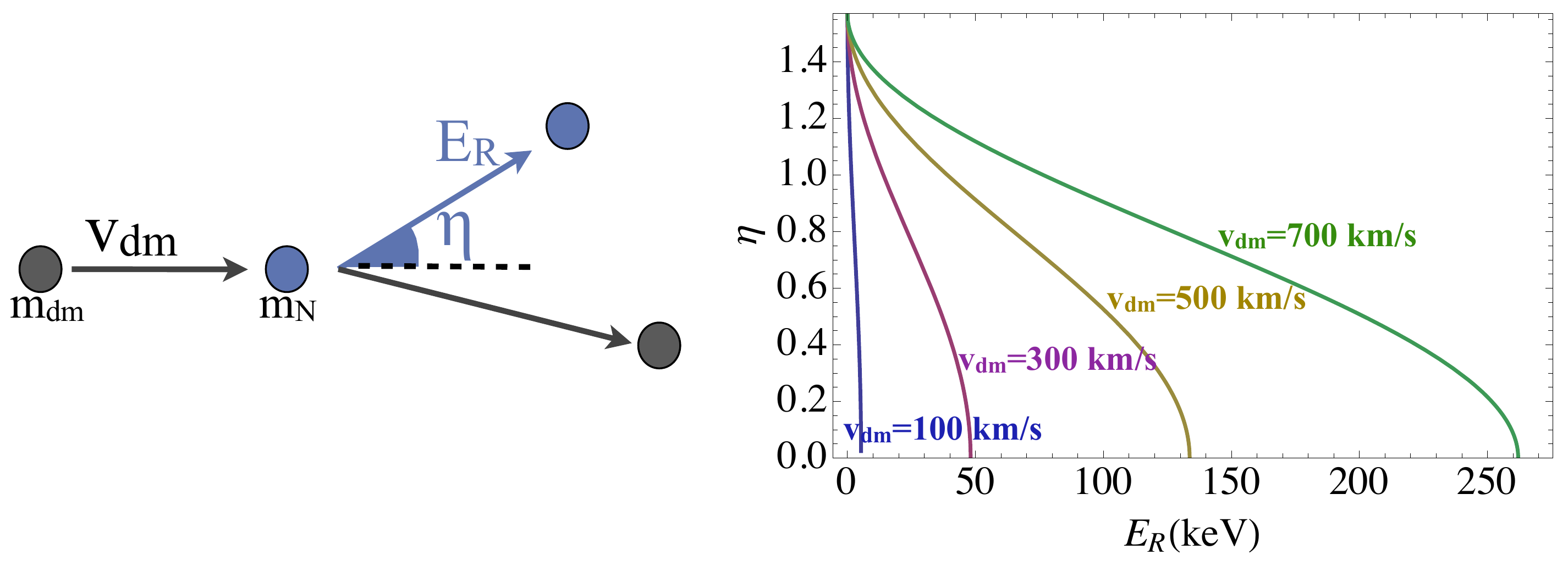}
\caption{Left: Kinematics of the DM-nucleus scattering. Right: The curves illustrate the relationship in Eq.~\ref{eqkine} for different values of the initial DM velocity $\vec{v}_\text{dm}$, assuming $m_\text{dm}=100\GeV$ and the scattering is elastic off of a Germanium target.}
\label{kine}
\end{figure}
Above, $v_\text{min}(E_R)$ is the minimum dark matter velocity required by energy and momentum conservation in order for the nucleus to be scattered with recoil energy $E_R$. If the kinematics of the scattering is elastic, $v_\text{min}(E_R)$ is given by
\begin{equation}
v_\text{min}(E_R)=\sqrt{\frac{m_N E_R}{2 \mu^2}},
\end{equation}
where $\mu$ is the reduced mass of the dark matter-nucleus system.  Fig.~\ref{kine} illustrates the relation in Eq.~\ref{eqkine} for different values of the initial dark matter velocity.

\begin{figure}[t]
\includegraphics[width=6.5in]{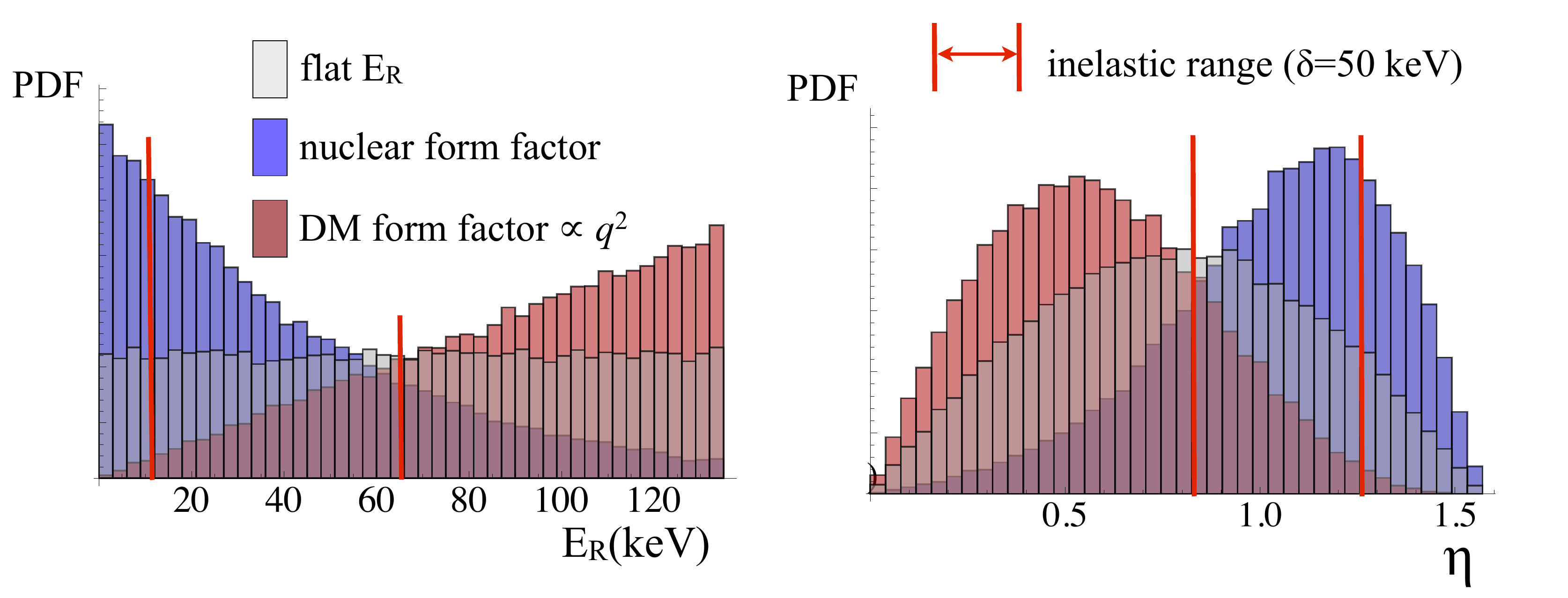}
\caption{Recoil energy (left) and recoil angle (right) distributions for a fixed dark matter velocity $\vec{v}_\text{dm}$. A flat scattering probability over the kinematically allowed energy range will translate into a nuclear recoil directional pdf peaking at an angle of $\pi/4$ with respect to the incident dark matter direction. Form factors will introduce an energy dependence of the scattering probability, distorting not only the recoil spectrum but also the nuclear directional rate. Kinematic properties such as inelasticity will not distort the pdf's, but truncate them over the kinematically allowed range.}
\label{dists}
\end{figure}

Notice that given $\vec{v}_\text{dm}$, the angle $\eta$ and the nuclear recoil energy $E_R$ are not uniquely determined. In fact, given $\vec{v}_\text{dm}$, the recoil energy can range in the interval
\begin{equation}
\label{ERrange}
0\leq E_R\leq\frac{2 \mu^2 v_\text{dm}^2}{m_N},
\end{equation}
for elastic scattering. Given a flux of dark matter particles over the detector, the nuclear recoil rate is determined by Eq.~\ref{eqkine} convolved with the dark matter velocity distribution and the scattering probability density over the energy recoil range in Eq.~\ref{ERrange}. If this probability distribution function (pdf) is flat in $E_R$, the corresponding probability distribution over $\eta$ will range from 0 to $\pi/2$, peaking at $\pi/4$, as illustrated in Fig.~\ref{dists}.

However, generically the scattering cross section will have a dependence on energy, modifying the probability that, given $\vec{v}_\text{dm}$, the nucleus will recoil with energy $E_R$. For instance, the finite size of the nucleus will introduce a nuclear form factor dependence that suppresses the scattering probability at high recoil energies \cite{Lewin:1995rx}. There may also be a form factor associated with the nuclear spin dependence of the interaction \cite{Feldstein:2009tr,Chang:2010en,Barger:2010gv}, or on the dark matter properties, for instance if the dark matter has a dipole moment or is a composite state \cite{Nussinov:1985xr,Chivukula:1989qb,Bagnasco:1993st,Foadi:2008qv,Gudnason:2006yj,Khlopov:2008ki, Alves:2009nf,Fitzpatrick:2012ix}. Such form factors will alter not only the nuclear recoil energy distribution, but the nuclear recoil direction as well.

Fig.~\ref{dists} illustrates how form factors distort the distributions on $E_R$ and $\eta$ for a fixed $\vec{v}_\text{dm}$. For instance, the spin-independent nuclear form factor that parameterizes the loss of coherence in the scattering with the nucleus suppresses the probability of scattering at large recoil energies. The effect of that on directionality is to suppress the rate towards small $\eta$. A dark matter form factor depending on a positive power of the momentum transfer, such as $q^2$, will have the opposite effect, suppressing the scattering probability at low energies and shifting the recoil direction rate towards small angles with respect to the dark matter incident direction.

Another interesting possibility is that the scattering is inelastic, and the dark matter particle up-scatters to an excited state with mass splitting $\delta$ relative to the ground state \cite{TuckerSmith:2001hy}. In this case $v_\text{min}(E_R)$ will be given by
\begin{equation}
v_\text{min}(E_R)=\frac{1}{\sqrt{2 m_N E_R}}\left(\frac{m_N E_R}{\mu_N}+\delta\right).
\end{equation}
The presence of inelasticity will not alter the pdf's shape in $E_R$ and $\eta$ for a given $\vec{v}_\text{dm}$, but only truncate the pdf to a narrower range $[{E_R}_\text{min} , {E_R}_\text{max}]$ (and corresponding $[\eta_\text{min} , \eta_\text{max}]$) that satisfies the relation $v_\text{dm}=v_\text{min}({E_R})|_{{E_R}_\text{min},{E_R}_\text{max}}$.

\subsection{Directional dependence from the velocity distribution}
\label{Subsec::Anisotropies}

As mentioned in the previous subsection, the energy dependence of the dark matter - nucleus interaction affects the directional rate of nuclear recoils. In this section we discuss the additional effect of anisotropies in the velocity distribution.

So far we considered the effect of dark matter incidence with fixed velocity and direction. That scenario may be realized if the DM distribution in the solar neighborhood is dominated by a stream with low velocity dispersion \cite{Lisanti:2011as,Kuhlen:2012fz,Vergados:2012xn}. Another possibility is that the dominant dark matter component in the solar neighborhood is isotropic and at equilibrium. As a first approximation we can consider a thermal velocity distribution truncated at some escape velocity, $v_\text{esc}$. If the dark matter halo and the solar system were co-rotating, this distribution would yield an isotropic nuclear recoil rate. However, if the solar system is moving through a static halo, the flux of dark matter particles will be boosted in the direction of the Earth's motion, creating interesting effects such as an annual modulation in the rate and a large anisotropy in the nuclear recoil direction, which will peak in the opposite direction to the Cygnus constellation. Anisotropic velocity distributions may generate even more interesting features in the directional rate of nuclear recoils, such as rings \cite{Bozorgnia:2011vc} and hot spots.

\begin{figure}[t]
\includegraphics[scale=0.4]{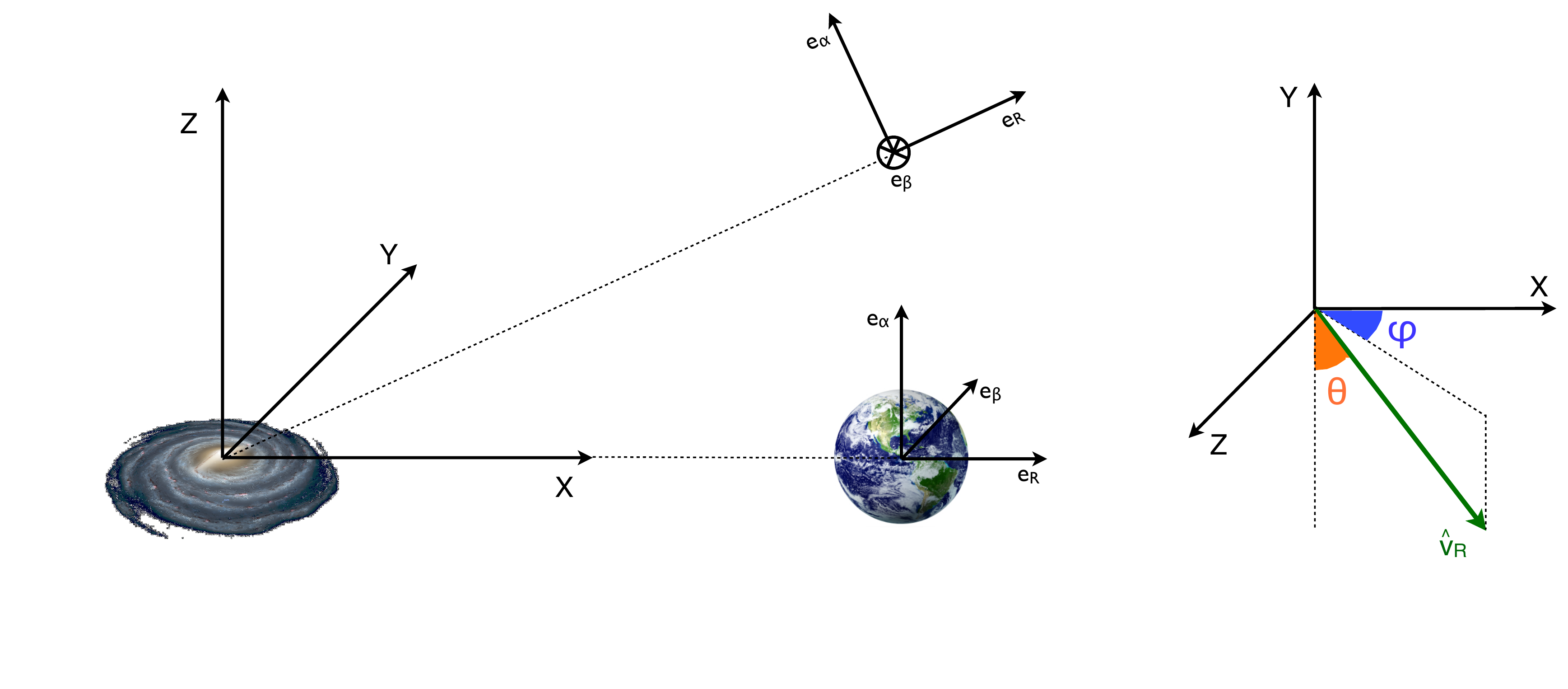}
\caption{(Left) The coordinate system adopted in this study. The $x$-direction points away from the center of the halo, and the $z$-direction is aligned with the axis of the disk. (Right) Angles $\theta$ and $\phi$ parameterizing the nuclear recoil direction: $\theta$ is the angle between the $\vec{v}_{R}$ and $-\vec{v}_{\oplus}$\label{LSR}, and $\phi$ is the azimuthal angle with respect to the Earth's motion.}
\end{figure}

In order to illustrate those features, we begin by establishing formulas and conventions that will be used throughout this paper. Given a dark matter velocity distribution $f(\vec{v})$ in the galactic frame, the nuclear recoil rate as a function of energy and direction is given by \cite{Gondolo:2002np,Alenazi:2007sy,Vergados:2012ze,Vergados:2001rr}:
\begin{eqnarray}
\label{Eq: Rate}
\frac{dR}{dE_Rd\Omega_R}=\mathcal{N}~F(E_R) \int~ d^3\vec{v} ~f(\vec{v})~\delta\left[(\vec{v}-\vec{v}_{\oplus})\cdot\hat{v}_R-\vmin(E_R)\right]
%\Theta[v_{\esc}^2 - v^2],
\end{eqnarray}
where $\vec{v}_{\oplus}$ is the Earth's velocity in the galactic frame. $\mathcal{N}$ contains the dependence on the scattering cross-section at zero momentum-transfer, on the local dark matter density and on phase-space factors. $F(E_R)$ parameterizes the cross-section at finite momentum transfer as a product of nuclear and dark matter form factors. Finally, $\hat{v}_R$ is the direction of the nuclear recoil. We will ignore the Earth's orbit around the Sun and choose coordinates such that
\begin{eqnarray}
\vec{v}_{\oplus}&\equiv&(~0~,~v_{\oplus}~,~0~),\\
\hat{v}_r&\equiv&(~\sin\theta\cos\phi~,~ \cos\theta~,~ \sin\theta\sin\phi~).
\end{eqnarray}

In those coordinates the $z$-direction is aligned with the axis of the disk and the $x$-direction points away from the center of the galaxy as depicted in Fig.~\ref{LSR}. We will also neglect the eccentricity of the Sun's orbit around the center of the galaxy. In that limit the Earth's velocity is parallel to the $y$-direction.
%The coordinate system used here is the one corresponding to the LSR frame, with the $z$-direction being aligned with the axis of the disk and the $x%-direction pointing away from the center of the galaxy (radial direction). In this study, we substract the motion of the Earth around the Sun and neglect the eccentricities of the Sun orbit around the center of the galaxy so that the velocity of the Earth is approximated to be parallel to the $y%-direction.

Note from Eq.~\ref{Eq: Rate} that, if the velocity distribution $f(\vec{v})$ is isotropic in the galactic frame, then the only source of anisotropy in the rate will come from the Earth's boost, $\vec{v}_{\oplus}\cdot\hat{v}_R=v_{\oplus}\cos\theta$. Therefore, in the coordinate system we adopted, isotropic velocity distributions should induce rates that have a dependence on $\theta$ but not on $\phi$. That means that any $\phi$ dependence on the rate is a smoking gun signature of an anisotropic velocity distribution.

\begin{figure}[t]
\centering
\includegraphics[width=6.0in]{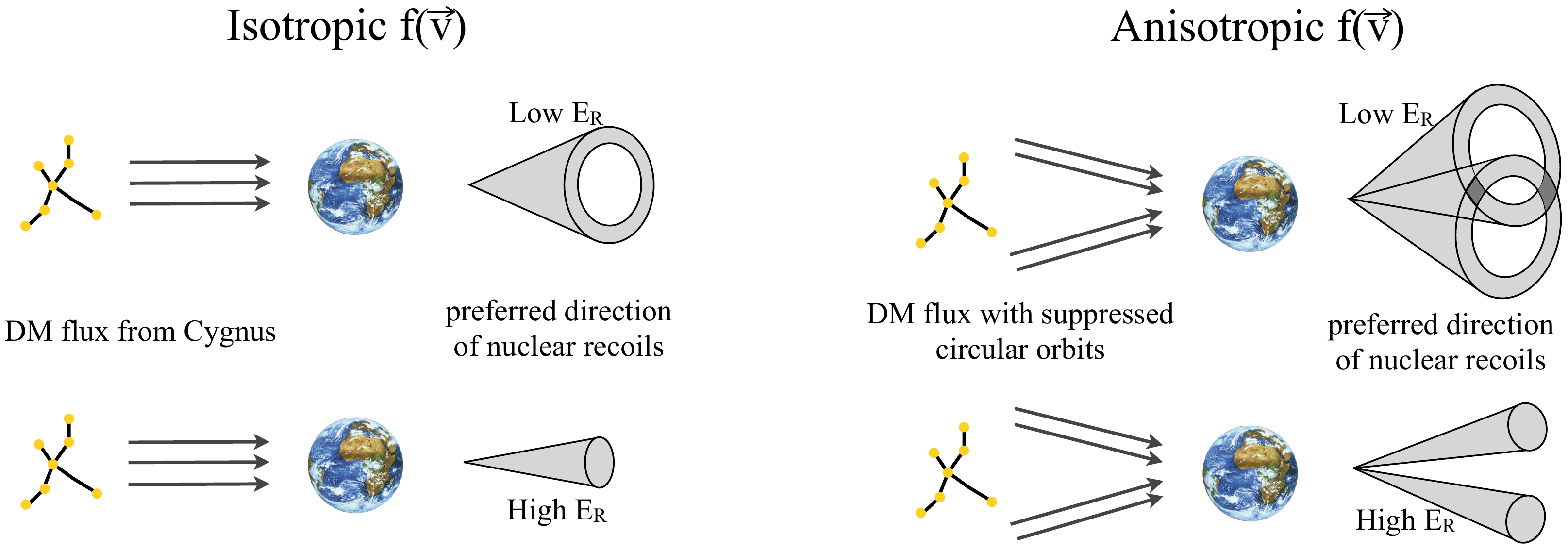}
\caption{Nuclear recoil directions for dark matter particles scattering at high velocities (up) and low velocities (down) for isotropic (left) and anisotropic (right) dark matter velocity distributions\label{lowv}}
\end{figure}

\begin{figure}[h]
\centering
\includegraphics[width=6.5in]{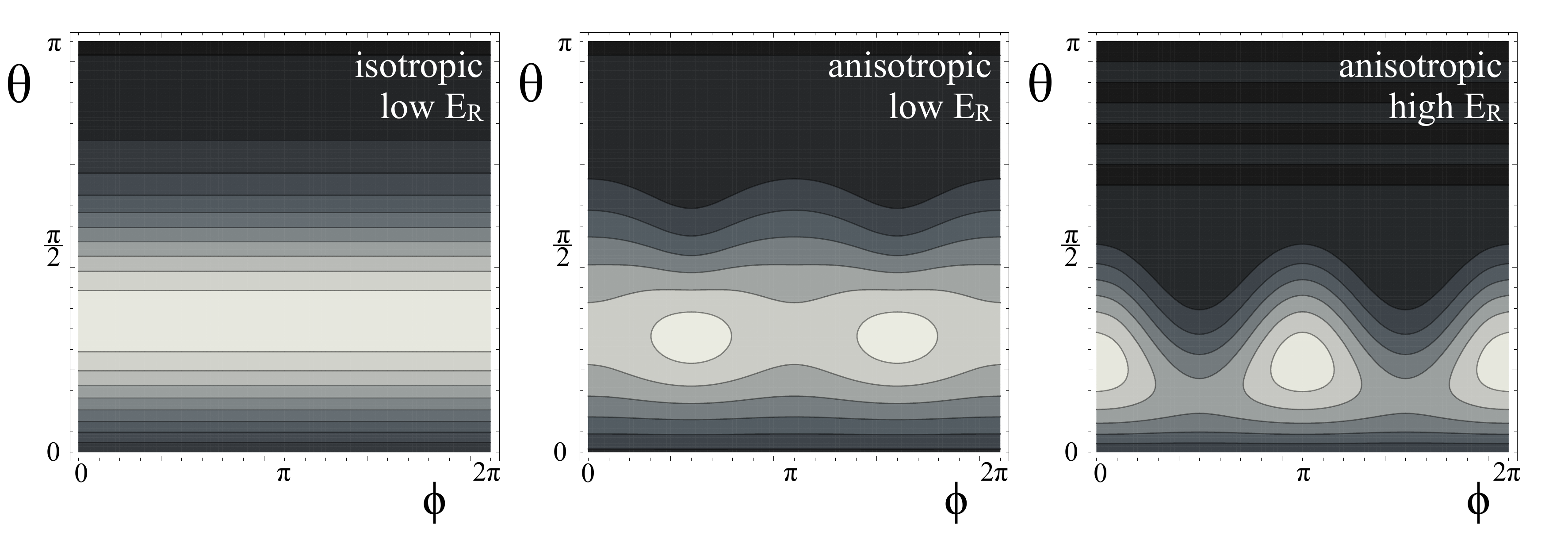}
\caption{$\theta-\phi$ profiles of the directional detection rates for an isotropic dark matter velocity distribution at low recoil energy (left) and an anisotropic dark matter velocity distribution at low (middle) and high (right) recoil energies, where the anisotropy originates from suppressed circular orbits. \label{Fig: ThetaPhiProfiles}}
\end{figure}

For illustration purposes, we compare the angular dependence of the nuclear recoil rate for an isotropic versus an anisotropic distribution. As pointed out, the predicted rate for the isotropic distribution is flat on $\phi$, and has a dependence on $\theta$ that varies with the recoil energy. If the energy is low, the recoil direction will prefer high values of $\theta$, given Eq.~\ref{eqkine} and the fact that $v_\text{min}(E_R)$ decreases as $E_R$ decreases. On the other hand, increasing the recoil energy will shift the rate towards lower values of $\theta$. That is illustrated pictorially in Fig.~\ref{lowv} - low recoil energies exhibit ``cone-like'' features in the recoil direction that point away from the incoming dark matter flux. Increasing the recoil energy has the effect of collimating the ``recoil cone'', until the ring-like structure finally shrinks to a hot spot.

Now consider an anisotropic velocity distribution for which the tangential components of the velocities are suppressed relative to the radial component. In that case circular orbits are subdominant, and the preferred flux of dark matter particles is along the radial direction in the galactic frame. In the Earth's frame, however, the flux picks up a component along the Earth's motion. That generates two overlapping ``recoil cones'' and hot spots in the recoil rate. As in the isotropic case, increasing the recoil energy will shrink the recoil cones to two more pronounced hot spots that will be shifted in $\phi$ by $\pi/2$ relative to the low energy hot spots. That effect is illustrated in Fig.\ref{Fig: ThetaPhiProfiles}.

%s\vspace{5in}

\section{Integrals of Motion and Jeans Theorem}
\label{Sec: Jeans}

In the previous section we discussed how the particle and astrophysical properties of dark matter manifest themselves in directional nuclear recoil signatures. Since very little is known about those properties, educated guesses have to made in order to explore direct detection signals. However, in the event that a dark matter signal is discovered in direct detection, we would like to take the data at face value and interpret it free of theoretical prejudice whenever possible.

In this section we will develop the tools for the main goal of this work: to infer the dark matter phase space distribution in the solar neighborhood from directional data. The first step towards this goal is to choose the correct variables on which to express the phase space distribution function. Inquiring how the dark matter halo was formed in the first place will give us some guidance. For instance, since the Milky Way halo has not undergone recent mergers, we can assume that the majority of dark matter particles has had enough time to relax to a state of equilibrium, in which case Jeans Theorem should hold to a good approximation \cite{binney1987galactic}:

\textbf{Jeans theorem} :~{\it Any steady-state solution of the collisionless Boltzmann equation depends on the phase-space coordinates only through integrals of motion. Conversely, any function of the integrals of motion is a
steady-state solution of the collisionless Boltzmann equation.}

%\subsubsection{Jeans theorem}

%Directional detection experiments would give a unique insight to the structure of the dark matter velocity distribution function, allowing especially to identify a wide class of anisotropic distributions. Nevertheless, the  dark matter velocity distribution function measured on Earth is different from the velocity distribution function at other places in the galaxy. For haloes at equilibrium though, a fundamental law governing the behavior of a large class of particle systems, Jeans theorem, allows to deduce the global dark matter halo distribution function from the local dark matter velocity distribution \cite{binney1987galactic}.

%\textbf{Strong Jeans theorem :} \emph{The distribution function of a stead$y%-state system in which almost all orbits are regular with non-resonant frequencies may be presumed to be a function only of three independent isolating integrals, which may be taken to be the actions.}

%For any dark matter halo at equilibrium having $n$ independent integrals of motion $I_1\ldots I_n$, the associated DF can always be expressed in function of at most three of them, for example, $I_1$, $I_2$ and $I_3$.

The Jeans theorem is the inspiration for the parameterization of the dark matter DF that will be used in this work. If a measurement of the local dark matter distribution function were our sole concern, any parameterization in terms of functions of the phase space coordinates could be adopted, even if such functions were not the true integrals of motion. But in scenarios where the equilibrium conditions of Jeans Theorem are valid, expressing the DF in terms of the true integrals of motion provides information beyond the scales of the solar neighborhood. In principle, we could analytically extend a measurement of the phase distribution in the solar system to any other position in the halo, and therefore infer the shape of the global DF through
\begin{eqnarray}
f(\vec r,\vec v) = f(I_1,I_2,..., I_n) = f(I_1(\vec r, \vec v),I_2(\vec r, \vec v),..,I_n(\vec r, \vec v)),
\end{eqnarray}
$I_1\ldots I_n$ being integrals of motion of the halo. % CHANGED : this sentence has been added

Moreover, the strong version of Jeans Theorem allows us to assume that, for all practical purposes, the phase-space distribution will be a function of at most 3 integrals of motion. Quasi-static haloes, for which Jeans theorem applies, always have energy, $\mathcal{E}$, as an integral of motion. Other possibilities include the components of the angular momentum, $\vec{L}$, and possibly non-classical integrals which have no analytical expression. Interestingly, the directional recoil rate is easily integrated if the distribution function depends only on energy. Consider a generic DF $f(\mathcal{E})$. According to Eq.\ref{Eq: Rate}, the directional rate is given by 
\begin{eqnarray}
\frac{dR}{dE_Rd\Omega_R}\propto\int~\delta\left(\vec{v}\cdot\hat{v}_R-\rho\right)~ d^3v ~f\left(\mathcal{E}\right)~\Theta(-\mathcal{E}),
\end{eqnarray}
where we have defined
\begin{eqnarray}
\rho\equiv\rho(E_R,\theta,\phi)\equiv\vec{v}_{\oplus}\cdot\hat{v}_R+\vmin(E_R),
\end{eqnarray}
and the energy $\mathcal{E}$ is given in terms of the dark matter's velocity and the local galactic escape velocity $v_\text{esc}$:
\begin{equation}
\mathcal{E}\equiv\frac{v^2}{2}-\psi(\vec r_{\oplus})=\frac{v^2-v^2_{\text{esc}}}{2}.
\end{equation}
The factor $\Theta(-\mathcal{E})$ excludes the possibility that the there are unbound particles (for which $\mathcal{E}>0$).

On Appendix A we show that, 
\begin{eqnarray}
\label{RateEton}
\int~\delta\left(\vec{v}\cdot\hat{v}_R-\rho\right)~ d^3v~\Theta(-\mathcal{E}) ~\mathcal{E}^n=2\pi ~\frac{(-1)^n}{n+1}\left(\frac{v^2_{\text{esc}}-\rho^2}{2}\right)^{n+1}~\Theta\left(v^2_{\text{esc}}-\rho^2\right).
\end{eqnarray}

Hence, expanding $f(\mathcal{E})$ in a Taylor series, using Eq.~\ref{RateEton} and re-summing the expansion we finally obtain:
\begin{eqnarray}
    \label{remarkable}
\frac{dR}{dE_Rd\Omega_R}&\propto&\int~\delta\left(\vec{v}\cdot\hat{v}_R-\rho\right)~ d^3v ~f\left(\mathcal{E}\right)~\Theta(-\mathcal{E})\\
&\propto&2\pi~\Theta\left(v^2_{\text{esc}}-\rho^2\right)~\int_{-\left(\frac{v^2_{\text{esc}}-\rho^2}{2}\right)}^{0}~f(\mathcal{E})~d\mathcal{E}.
\end{eqnarray}

Eq.~\ref{remarkable} is a remarkable relation between the directional nuclear recoil rate and the underlying dark matter distribution. It makes it clear that, if the DF is isotropic in the solar neighborhood, measuring the dependence of the rate on $v^2_{\text{esc}}/2-\rho(E_R,\theta,\phi)^2/2$ is equivalent to directly measuring the underlying DF, since the later is the derivative of the rate with respect to $v^2_{\text{esc}}/2-\rho(E_R,\theta,\phi)^2/2$.

%The most common integrals of motion for stellar systems are the energy $E$, the components of the angular momentum $\vec L$, and two non-classical integrals $I_2$ and $I_3$ which have no analytical expression. Quasi-static haloes, for which Jeans theorem applies, always have $E$ as an integral of motion and therefore, the DFs associated to most stationary halo models are of the form $f(E,\mathcal{I}_2,\mathcal{I}_3)$. In the rest of this paper, $\mathcal{I}_2$ and $\mathcal{I}_3$ are chosen to be two angular momentum components, respectively $L_z$ and $L_t=\sqrt{L^2-L_z^2}$. Note that near the Earth, $L_t=|L_y|$.
%Choosing angular momentum components as integrals of motion has no immediate justification though it is the only way to be able to compute the differential detection rate associated to a given dark matter distribution function. In section~\ref{Sec: VLII}, it actually proves to be a reasonable hypothesis near the center of the galaxy for a dark matter halo generated using the Via Lactea II N-body simulation.

More generically, however, the dark matter DF will be anisotropic and depend on other integrals of motion, for instance the magnitude of the angular momentum $|\vec{L}|$ or its components, such as $L_z$. Unfortunately, no straightforward expression for the rate has been derived for a generic function $f(\mathcal{E},|\vec{L}|,L_z)$. However, we have been able to analytically integrate the rate for a certain class of functions (see Appendix~\ref{appendix}). Such functions have polynomial dependence in $\mathcal{E}$ and harmonic dependence in the components of the angular momentum.
%CHANGED : removed separability since it's not necessary to have an analytical expression for the rate
That will prove useful for the method we will propose in the next section, where we will perform a special function decomposition of the DF.

\section{Special Function Decomposition of the Halo Distribution}
\label{decomposition}
Although the reconstruction of the dark matter distribution function from directional and non-directional detection signals has already been investigated for specific halo types \cite{Green:2010gw,Peter:2011wi,Peter:2009ak,Drees:2007hr,Lee:2012pf}, no model independent reconstruction study has been performed yet (see, however, halo independent proposals to interpret direct detection signals \cite{Fox:2010bz,Gondolo:2012rs,Frandsen:2011gi}). One of the main goals of this paper is to provide a parameterization of the DF that is as generic as possible. The method we propose is to decompose the distribution function in an orthonormal basis of special functions, and fit the measured rate to the coefficients of this decomposition. If the basis is appropriately chosen, the higher order coefficients will be subdominant and we will be able to truncate the series to a finite and preferably  manageable number of coefficients.

There is no optimal choice for the variables and for the basis of decomposition that applies for all possible cases, and our main purpose is to illustrate out method rather than explore all possible variables and bases. Therefore, in this paper we will choose the following integrals of motion on which to parameterize the local dark matter distribution function: 

\begin{itemize}
{\item The energy $\mathcal{E}$, which, as before, is given by $\mathcal{E}=\frac{v^2-v^2_{\text{esc}}}{2}$}.
{\item The $z$-component of the angular momentum, $L_z$, which in the coordinate system we have adopted is given by $L_z=\sqrt{r^2-z^2}v_\phi$}. In the solar neighborhood $z=0$ and hence $L_z=r_\oplus v_\phi$.
{\item The magnitude of the angular momentum, whose DF dependence will be incorporated through the variable $L_t\equiv\pm\sqrt{|\vec{L}|^2-L_z^2}=\sqrt{r^2v_\theta^2+z^2v_\phi^2}$. Again, in the solar neighborhood that reduces to $L_t=r_\oplus v_\theta$.}
\end{itemize}

We will make a further assumption which is not generic but that will greatly facilitate the computation of the rate and reduce the number of decomposition coefficients in the fit. This assumption is that the DF $f(\mathcal{E},L_t,L_z)$ is separable:
\begin{eqnarray}
\label{separable}
f(\mathcal{E},L_t,L_z)& =f_1(\mathcal{E})f_2(L_t)f_3(L_z).
\end{eqnarray}

Under this assumption, we will adopt the following choice of basis for $f_1(\mathcal{E})$, $f_2(L_t)$ and $f_3(L_z)$:
\begin{eqnarray}
\label{expansion}
f_1(\mathcal{E})&=&\sum_{\ell}c_{P_\ell}\tilde{P}_\ell\left(\frac{\mathcal{E}}{\mathcal{E}_\text{lim}}\right),\nonumber\\
f_2(L_t)&=&\sum_{n}c^t_{F_n}\cos\left(n\pi \frac{L_t}{L_{\text{max}}}\right),\\
f_3(L_z)&=&\sum_{m}c^z_{F_m}\cos\left(m\pi \frac{L_z}{L_{\text{max}}}\right)\nonumber.
\end{eqnarray}

Above, $f_1(\mathcal{E})$ is expanded in shifted Legendre polinomials, $\tilde{P}_\ell$, and $\mathcal{E}_\text{lim}$ is the lowest dark matter energy that an experiment with recoil threshold $E_R^\text{th}$ is able to probe:
\begin{eqnarray}
\mathcal{E}_\text{lim}\equiv\frac{(v_\text{min}(E_R^\text{th})-v_\oplus)^2-v^2_{\text{esc}}}{2}.
\end{eqnarray}
In principle we could have chosen $\mathcal{E}_\text{lim}$ to the lowest possible energy, $-v^2_\text{esc}/2$. However, since a realistic experiment will have no information about particles with energy less than $\mathcal{E}_\text{lim}$, there is no purpose in parameterizing the distribution function beyond what is measurable.
The functions $f_2(L_t)$ and $f_3(L_y)$ are decomposed in a Fourier series, where $L_\text{max}\equiv r_\oplus v_\text{esc}$.

A great practical advantage of the parameterization above is that the recoil rate can be analytically integrated. That way no numerical integration is necessary, reducing computational time to perform the fits. In Appendix~\ref{appendix} we derive and list the analytical forms for the rate on those basis functions.

In the following section we will illustrate how the fit is performed for a hypothetical experiment measuring the recoil rate for an anisotropic DF. We will choose as an example a Michie distribution \cite{Michie:1963a} and generate mock up Monte Carlo data as the detector would measure it. We will then employ our method to recover the original distribution. In Sec.~\ref{Sec: VLII} we will use the Via Lactea II simulation as the underlying dark matter distribution and simulate the rate at the detector. We will then employ our method to perform the fits.

\section{Directional Fits to a Michie Distribution }
\label{Sec: VDF}

In this section we illustrate our method for a hypothetical dark matter signal in a directional detector. Ideally we would like to have a large number of events to perform the fit. However, the current limits are quite stringent for spin-independent elastic scattering dark matter particles with mass of $\OO(100)$ GeV. For instance, the latest XENON100 run with 4 kg-yrs of exposure observed 3 candidate events with an expected background of $1.8\pm0.6$ events \cite{Aprile:2011hi}. That suggests that, in the heavy WIMP scenario where discovery is just around the corner, the detection of $\OO(100)$ events would require at least hundreds of kg-yrs of exposure, which is still an unrealistic goal for the $\OO(1)$-kg detectors planned by current directional experiments. The most optimistic possibility is if dark matter is light, of order of a few GeV. In this mass range the limits are much less stringent \cite{Angle:2011th,Ahmed:2010wy}, and the relatively light targets considered by directional experiments, such as $\text{CF}_4$ and $\text{CS}_2$, are ideal.

Therefore we choose to generate this hypothetical signal for an elastically scattering dark matter particle with mass $m_\text{dm}=6\GeV$. We assume that the scattering is off of a $\text{CS}_2$ target, with recoil energy acceptance window  $E_R \in [5 \keV,15 \keV]$ and for simplicity we ignore the scattering off of carbon. We assume that the experiment operates in a zero background environment, but in principle background events could be straightforwardly incorporated in the fits if their directional shape is known. Finally, the results we will show in the following are obtained from fits without resolution effects taken into account. As a crude approximation of the detector angular resolution, we have divided the nuclear recoil direction in bins of size $(10^\circ,10^\circ)$ and did not observe any significant alteration of our results. In the results shown in the rest of this article, no binning has been used.

%In the limit of infinite statistics and infinite detector resolution, the dark matter distribution function could theoretically be unambiguously determined from any experimental signal. Studying how well the shape of the DF can be determined in function of the number of dark matter events detected will be of great importance for future directional detection experiments. In this purpose, mock dark matter directional detection signals of different sizes are generated and fitted. The evolution of the error on the DF in function of the size of the datasets is then studied.

%\subsection{Detector characteristics, dark matter properties and benchmark model}

%The dark matter particles are scattered elastically on ${}^{32}\text{S}$ nuclei in a detector with directional sensitivity, lower energy threshold $E_{\text{min}} = 5 \keV$ and higher threshold $E_{\text{max}} = 15 \keV$. The dark matter mass is  taken to be $m_{DM} = 6 \GeV$. This choice of parameters leads to a very high $v_{\low}$ ($\approx 518 \kms$) but the approach exposed here should be the same even for lower values of the minimum velocity.

In order to illustrate the interesting effects of anisotropy, we choose a Michie distribution for the dark matter. In terms of integrals of motion, the Michie distribution is given by:
\begin{eqnarray}
\label{Michie}
f_{\text{Michie}}(\mathcal{E},|\vec{L}|)=~\mathcal{N}~e^{-\alpha|\vec{L}|^2/L_0^2}~\left(e^{-2\mathcal{E}/v_0^2}  - 1\right)~\Theta(-\mathcal{E})
\end{eqnarray}
Our choice of parameters is the following: $v_0 = 280 \kms$, $v_{\esc} = 600\kms$, $L_0=r_\oplus v_0$ and $\alpha=1$. We will assume in the fits that local escape velocity $v_\text{esc}$ as well as the dark matter mass $m_\text{dm}$ are know to sufficient precision. That is a reasonable assumption since directional experiments will only come into play after the recoil spectrum has been well measured by several non-directional experiments with different nuclear targets.

Note that the Michie distribution (\ref{Michie}) trivially satisfies the  separability condition (\ref{separable}) when expressed in terms of $\mathcal{E}$, $L_t$ and $L_z$, since $|\vec{L}|^2=L^2_t+L^2_z$. Then, with
\begin{eqnarray}
f_1(\mathcal{E})&=&e^{-2\mathcal{E}/v_0^2}  - 1,\\
f_2(L_t)&=&e^{-\alpha L_t^2/L_0^2},\\
f_3(L_z)&=&e^{-\alpha L_z^2/L_0^2}
\end{eqnarray}
we can perform the special function decomposition as in Eq.\ref{expansion} and obtain the coefficients:
\begin{eqnarray}
\{c_{P0},c_{P1},c_{P2},c_{P3},~c_{P4},c_{P5},...\}&=&\{1,1.65, 0.890, 0.297, 0.0722, 0.0137,...\},\\
\{c^t_{F0},c^t_{F1},c^t_{F2},c^t_{F3},...\}&=&\{0.412, 0.485, 0.0952, 0.00773,...\},\\
\{c^z_{F0},c^z_{F1},c^z_{F2},c^z_{F3},...\}&=&\{0.412, 0.485, 0.0952, 0.00773,...\}.
\end{eqnarray}
The overall normalization of the coefficients above is unphysical since it can be reabsorbed in the cross section. Hence, the coefficients were scaled such that $f_1(\mathcal{E})$, $f_2(L_t)$ and $f_3(L_z)$ are normalized to unity. We have truncated the energy expansion in Legendre polynomials to the 5th order, and the Fourier decomposition of the components of the angular momentum to the 3rd harmonic. The truncation of the expansions at this orders approximates the true functions to an accuracy better than the percent level.

%The dark matter local velocity distribution is taken to be a Michie distribution with $\alpha=1$, $v_0 = 280 \kms$ and $v_{\esc} = 600\kms$. In terms of the energy and angular momentum variables $E$, $L_t$ and $L_z$ introduced in Sec.~\ref{Sec: Jeans} the associated DF is of the form
%%

%\subsection{Fitting the dark matter detection rate}
%\label{SubSecFitRate}

%The dark matter distribution function is approximated as 
%%
%\begin{eqnarray*}
%f(E,L_t,L_z) &=& f_1(E)f_2(L_t)f_3(L_z)\\
%f_1(E)& =& \sum_{n=0}^{N_E}a_n P_n(E)\\
%f_2(L_t)&=&\sum_{l=0}^{N_{L_t}}b_l\cos(l \pi L_t)\\
%f_3(L_z)&=&\sum_{m=0}^{N_{L_z}}c_m\cos(m \pi L_z)
%\end{eqnarray*}
%%
%where the $P_n$ are Legendre polynomials of degree $n$. In what follows,  $N_E = 5$ and $N_{L_t} =N_{L_z}=3$. These values have been determined by fitting directly the distribution function of the chosen benchmark model and changing them will not modify the method presented here.

Knowing the true values of the decomposition coefficients, we can compare them to the values obtained from fits of the rate. We generated several Monte Carlo data ensembles, varying the number of observed events. We then performed unbinned log-likelihood fits in the 11-dimensional parameter space of decomposition coefficients using the publicly available code $\mathtt{MultiNest~v2.10}$, a Bayesian inference tool that uses a nested sampling algorithm to scan the parameter space  \cite{Feroz:2007kg,Feroz:2008xx}.

%After having computed the differential rate associated to each of the expansion terms shown in Eq.~\ref{expansion} in function of the $a_n$, $b_l$, $c_m$, a likelihood fit of these parameters is performed using the MultiNest nested sampling algorithm \cite{Feroz:2007kg,Feroz:2008xx}. 

The 1-$\sigma$ range of the fitted coefficients can be extracted by approximating the log-likelihood function near its maximum by a quadratic function of the parameters of the fit:
\begin{eqnarray}
\LL(\vec C) \approx \LL_{\text{max}} - \frac{1}{2}\sum_i \frac{C_i^2}{\sigma_i^2},
\end{eqnarray}
where $\vec C$ is the vector representing a given set of coefficients and $\vec{\sigma}$ is their standard deviation.

\begin{figure}[t]
    \includegraphics[width=6.7in]{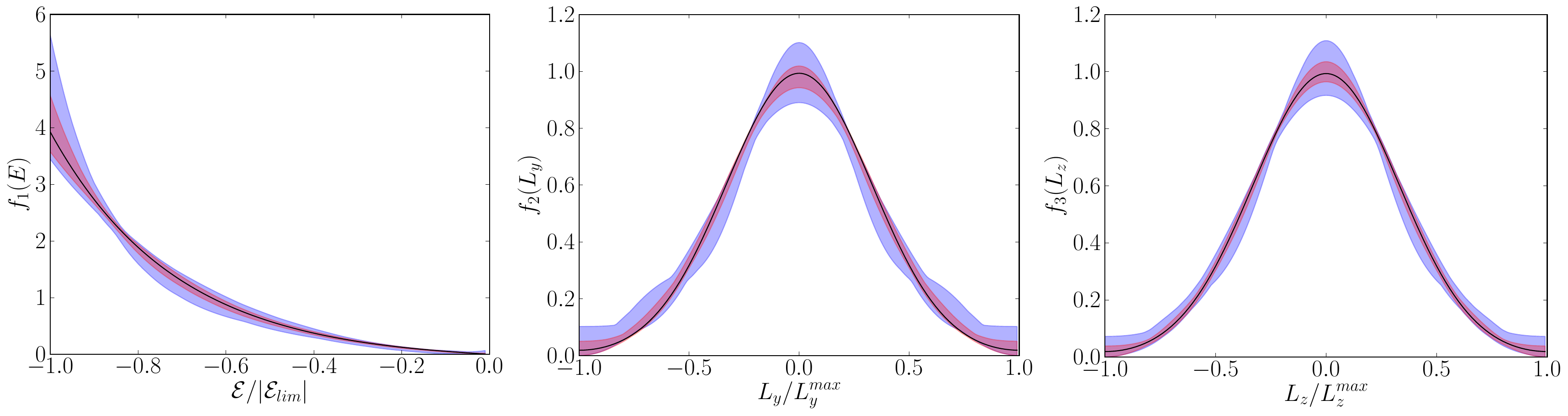}
    \caption{Results of the fits for $f_1(\mathcal{E})$, $f_2(L_t)$ and $f_3(L_z)$, assuming $10^4$ (red) and $10^3$ (blue) signal events. The bands delimit the ranges of these functions given by the 1-$\sigma$ errors on the decomposition coefficients. The underlying phase space distribution is given by the Michie DF in (\ref{Michie}), represented in graphs above by the solid lines.\label{bands}}
\end{figure}

\begin{figure}[h]
    \includegraphics[width=6.7in]{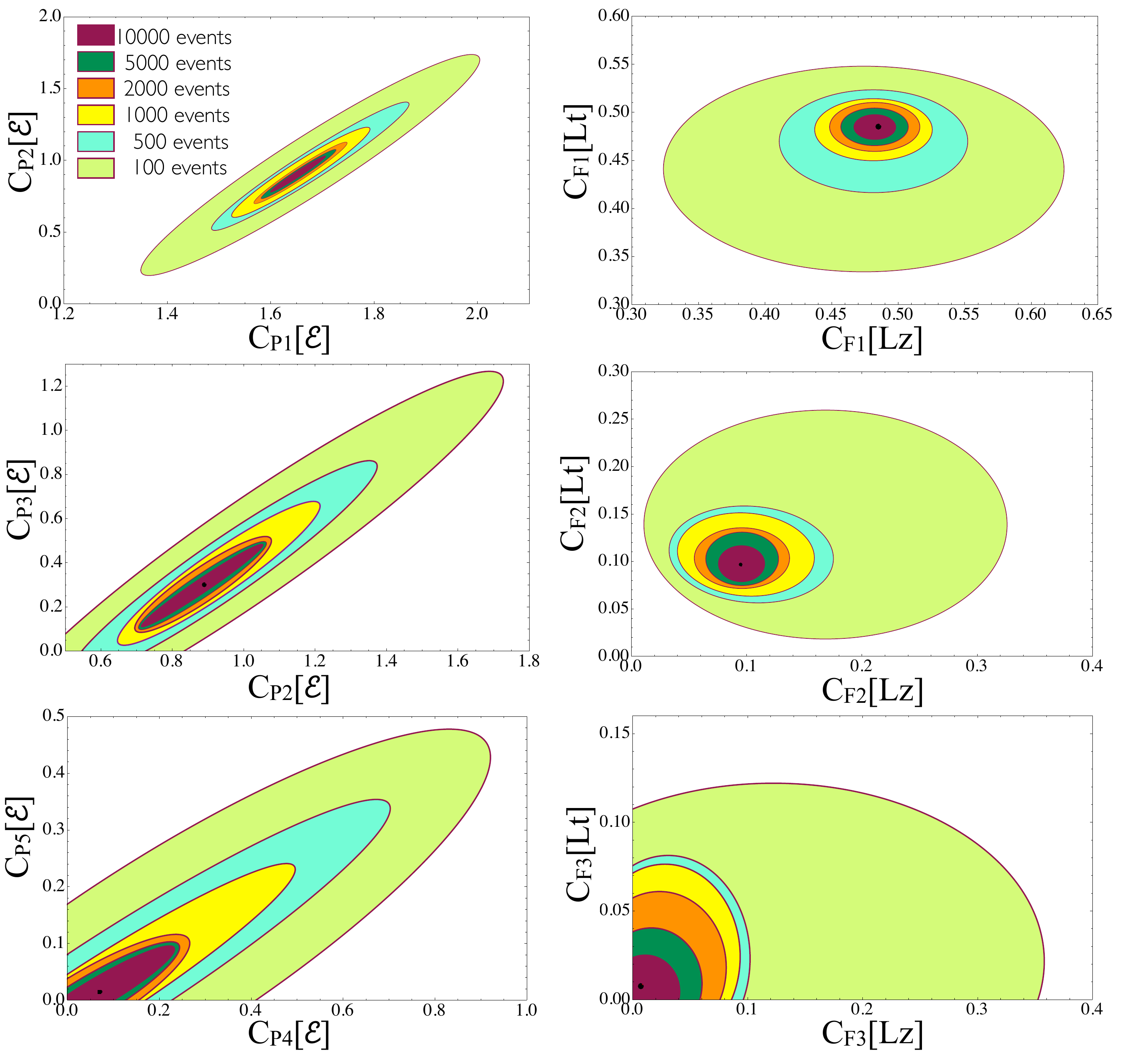}
    \caption{68\% confidence contours on the decomposition coefficients of the DM distribution, found by fitting several mock directional rates generated by Monte Carlo, with different numbers of observed events. The dots correspond to the true values of the coefficients.\label{ellipses}}
\end{figure}

In Fig.~\ref{ellipses} we show the results of the fit for different numbers of events, varying from 100 to 10,000. We display the 68\% C.L. contour intervals of the fitted coefficients, as well as the true values for comparison. As the errors are correlated, the 68\% range of the decomposition coefficients is actually a multidimensional ellipsoid. The different plots in Fig.~\ref{ellipses} are 2D projections of this ellipsoid in different planes of the parameter space.

As one can see, a few thousand events are necessary for a good measurement of the underlying dark matter distribution. Fig.~\ref{bands} displays the functional forms of $f_1(\mathcal{E})$, $f_2(L_t)$ and $f_3(L_z)$ resulting from fits with 1000 events and 10,000 events. This shows that our method works remarkably well if the variables and decomposition basis are chosen judiciously. In the real world, however, an appropriate choice of decomposition basis may not be as trivial as in this hypothetical example. However, several checks of the data can be made before performing fits to test whether a given hypothesis is a good approximation. We will illustrate that in the next section, where we will use the Via Lactea II N-body simulation as a template for the dark matter distribution in our galaxy, and ``scatter" the particles of this simulation is a detector placed at a location around 8 kpc from the center to try to infer the underlying phase space distribution.

\section{Fits to the Via Lactea II simulation}
\label{Sec: VLII}

No dark matter direct detection signal has been observed to this date to provide us with clues on the local distribution. Most of our current understanding of the range of possibilities for the local distribution comes from N-body simulations \cite{Kuhlen:2009vh,Lisanti:2011as,Kuhlen:2012fz,Lisanti:2010qx}, which also provide one of the best test grounds to investigate the formation and dynamics of dark matter halos \cite{Vogelsberger:2008qb}. With about $10^9$ particles with masses around $10^3$ solar masses, Via Lactea II is currently one of the highest resolution N-body simulations and can be used as a model of the Milk Way halo \cite{Kuhlen2008}.

In this section we will test the method proposed in this study using Via Lactea II as the underlying dark matter distribution. Ideally, we would like to choose a box located at $R_\oplus=8$~kpc away from the center of the halo (which is the distance of the solar system from the center of the Milky Way) and with a volume $(\Delta r)^3\ll R_\oplus^3$ where $\Delta r$ is the length of a side of the box. Unfortunately, in Via Lactea II a region of this size typically does not enclose any particles. Our approach is then to choose a spherical shell of 8kpc-radius and 200pc-thickness, which encloses enough particles to generate a dark matter signal at a directional detector. Since we do not want to wash out anisotropies, we choose a fixed position $\vec{r}_\oplus$ at which to place the detector, and for each dark matter particle in the shell with coordinates $\vec{v}$ and $\vec{r}$, we perform an affine map $\vec{v}(\vec{r})\rightarrow\vec{v}(\vec{r_\oplus})$.

%Besides, analyzing the dark matter haloes generated by these simulations would allow to make the experimental searches more efficient by narrowing down the parameter space.

%\subsection{Checking the validity of the hypotheses made about the dark matter distribution function}

In the same spirit of the previous section, we would like to fit a signal generated from Via Lactea II to a distribution function in terms of $\mathcal{E}$, $L_t$, $L_z$, which is separable as in Eq.~\ref{separable}, and decomposed as a Legendre series in $\mathcal{E}$ and a Fourier series in $L_t$, $L_z$, see Eq.~\ref{expansion}. However, before attempting to perform those fits we should test the validity of those assumptions.

\begin{figure}[t]
\includegraphics[scale=0.45]{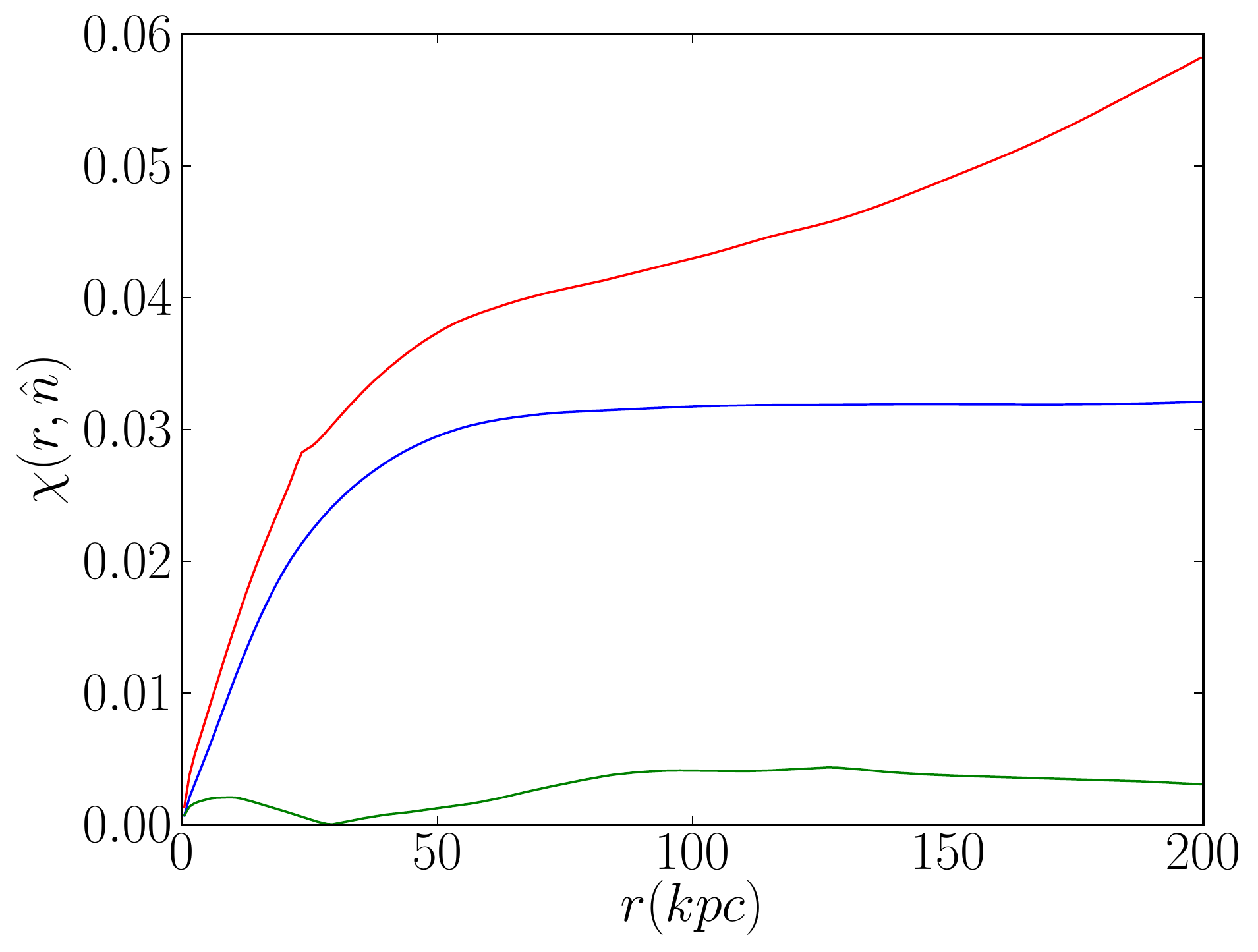}
\caption{The deviation of the gravitational potential from spherical symmetry, parameterized by $\chi(\vec{r})$, see Eq.~\ref{chi}. Three orthogonal directions for $\vec{r}$ are displayed: $\vec{r}=r\hat{x}$ (blue), $\vec{r}=r\hat{y}$ (green) and $\vec{r}=r\hat{z}$ (red). \label{psiXYZ} }
\end{figure}

The first of those assumptions, that the energy and angular momentum are integrals of motion, can be shown to be a good approximation for positions not too far away from the center of the halo, where 
%Previous studies of the VLII simulation have shown that the associated dark matter halo is triaxial. However, near the center of the galaxy,
the gravitational potential can be reasonably approximated as being spherical. A quantitative measure of the sphericity of the potential is given by
\begin{eqnarray}
\label{chi}
\chi(\vec{r}) =\left| \frac{\psi(\vec{r}) - \psi_s(r)}{\psi_s(\vec{r})}\right|.
\end{eqnarray}
Above, $\psi(\vec{r})$ is the exact gravitational potential at the position $\vec{r}$, and $\psi_s(r)$ is the potential computed assuming spherical symmetry and  evaluated using Gauss' law. $\chi(\vec{r})$ then parametrizes the deviation of the gravitational potential at position $\vec{r}$ from a spherically symmetric potential.

Fig.~\ref{psiXYZ} displays the $\vec{r}$ dependence of $\chi$ in three orthogonal directions: $\vec{r}=r\hat{x}$, $\vec{r}=r\hat{y}$ and $\vec{r}=r\hat{z}$. One can see that for $r<200 \kpc$, the assumption of a spherically symmetric potential leads to an error $\OO(6\%)$. For $r<30\kpc$, the error is reduced to $\OO(3\%)$. Hence the energy and angular momentum of dark matter particles within a $30\kpc$ radius should be approximately conserved.

%Expressing the dark matter velocity distribution function near the Earth in function of $E$, $L_t$ and $L_z$ would then allow to get a good approximation of the dark matter distribution function up to about $30 \kpc$ away from the center of the galaxy. Determining whether or not using these three variables provide a good description of the halo beyond $30 \kpc$ can be done only by comparing the dark matter velocity distributions at different places in the galaxy.

%\subsubsection{Separation of variables}

%Another important simplifying assumption about the dark matter distribution function is the hypothesis that $E$, $L_t$ and $L_z$ are separated, that is to say that the distribution function can be expressed as
%%
%\begin{eqnarray}
%\label{factor}
%f(E,L_t,L_z) = f_1(E)f_2(L_t)f_3(L_z)
%\end{eqnarray}
%%
The second of our assumptions, that of separability, can be evaluated through the function 
\begin{eqnarray}
G(\mathcal{E}, L_t, L_z) = \frac{g^2 f(\mathcal{E},L_t,L_z)}{g_\mathcal{E}(\mathcal{E})g_{L_t}(L_t)g_{L_z}(L_z)},
\end{eqnarray}
where
\begin{eqnarray}
g_\mathcal{E}(\mathcal{E}) &=& \int  f(\mathcal{E},L_t,L_z)dL_tdL_z\\
g_{L_t}(L_t) &=& \int  f(\mathcal{E},L_t,L_z)d\mathcal{E}dL_z\\
g_{L_z}(L_z) &=& \int  f(\mathcal{E},L_t,L_z)d\mathcal{E}dL_t\\
g &=& \int  f(\mathcal{E},L_t,L_z)d\mathcal{E}dL_tdL_z
\end{eqnarray}
which measures the correlation between the three variables $\mathcal{E}$, $L_t$ and $L_z$. If the dark matter distribution function verifies Eq.~\ref{separable}, $G(\mathcal{E}, L_t, L_z)$ should be constant and equal to $1$. 
\begin{figure}
\centering
\includegraphics[width=6.5in]{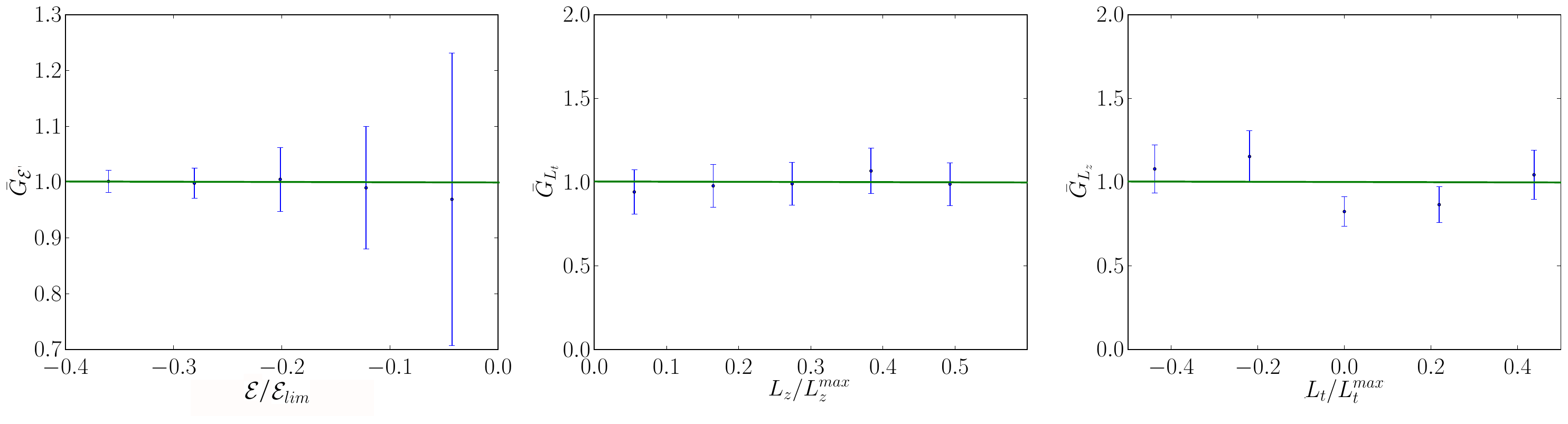}
\caption{$\bar G_\mathcal{E}$, $\bar G_{L_t}$ and $\bar G_{L_z}$ in function of $\mathcal{E}$, $L_t$ and $L_z$, respectively. \label{Corr}}
\end{figure}

Due to the lack of statistics, in Fig.~\ref{Corr} we plot $\bar G_\mathcal{E}$, $\bar G_{L_y}$, $\bar G_{L_z}$, the averages of $G$ over the two other variables, rather than the tridimensional behavior of the function. 
The fact that $\bar G_\mathcal{E}$, $\bar G_{L_t}$, $\bar G_{L_z}$ are consistent with a constant value equal to one reassures us that our assumption of separability is well justified. 

%Although $\bar G_{L_z}$ is more than one sigma away from $1$ for some values of $L_z$, it remains consistent with $1$ within two sigma and given the excellent behavior of $\bar G_\mathcal{E}$ and $\bar G_{L_t}$, separating the variables $\mathcal{E}$, $L_t$ and $L_z$ seems to be a well justified assumption. 

%\subsection{Results}

%\subsubsection{Detector and dark matter model}
\subsection{Local Fits to Via Lactea II}

The range of validity of our hypotheses described in Sec.~\ref{Sec: Jeans} were quantified in the previous section. In this section we perform fits to a directional detection signal generated from the distribution of VLII simulation, using particles contained within a shell located 8 kpc away from the center of the simulation, and with thickness of 200 pc. As mentioned previously, we fix the location for the detector at $\vec{r}_\oplus$, and transform the velocities of all particles in the shell through an affine map $\vec{v}(\vec{r})\rightarrow\vec{v}(\vec{r_\oplus})$.

As in Sec~\ref{Sec: VDF}, we set the mass of the dark matter particles to $m_\text{dm}=6\GeV$. Inspection of the VLII distributions allows us determine that the escape velocity at 8 kpc is $v_\text{esc}\approx545$~km/s.  We generate 10,000 events by elastically scattering the particles in the shell off of a $\text{CS}_2$ target, but ignore the scattering off of carbon. However, differently from Sec~\ref{Sec: VDF}, we are forced to lower the energy threshold of 5 keV. The reason for that is that with a 5 keV threshold, only particles in the tail of the energy distribution ($v\gsim518$~km/s) are kinematically  allowed to scatter. Unfortunately there only about $10^5$ particles inside the shell, and the tails of the distributions of those particles are not well populated. Hence, in order to avoid large errors  resulting from fitting the tails, we lower the energy threshold to 2 keV, allowing particles with $v\gsim330$~km/s to scatter. In that range the distribution is better populated and the fits are more reliable. As in Sec~\ref{Sec: VDF}, detector resolution effects are not taken into account.

\begin{figure}[h]
\includegraphics[width=4in]{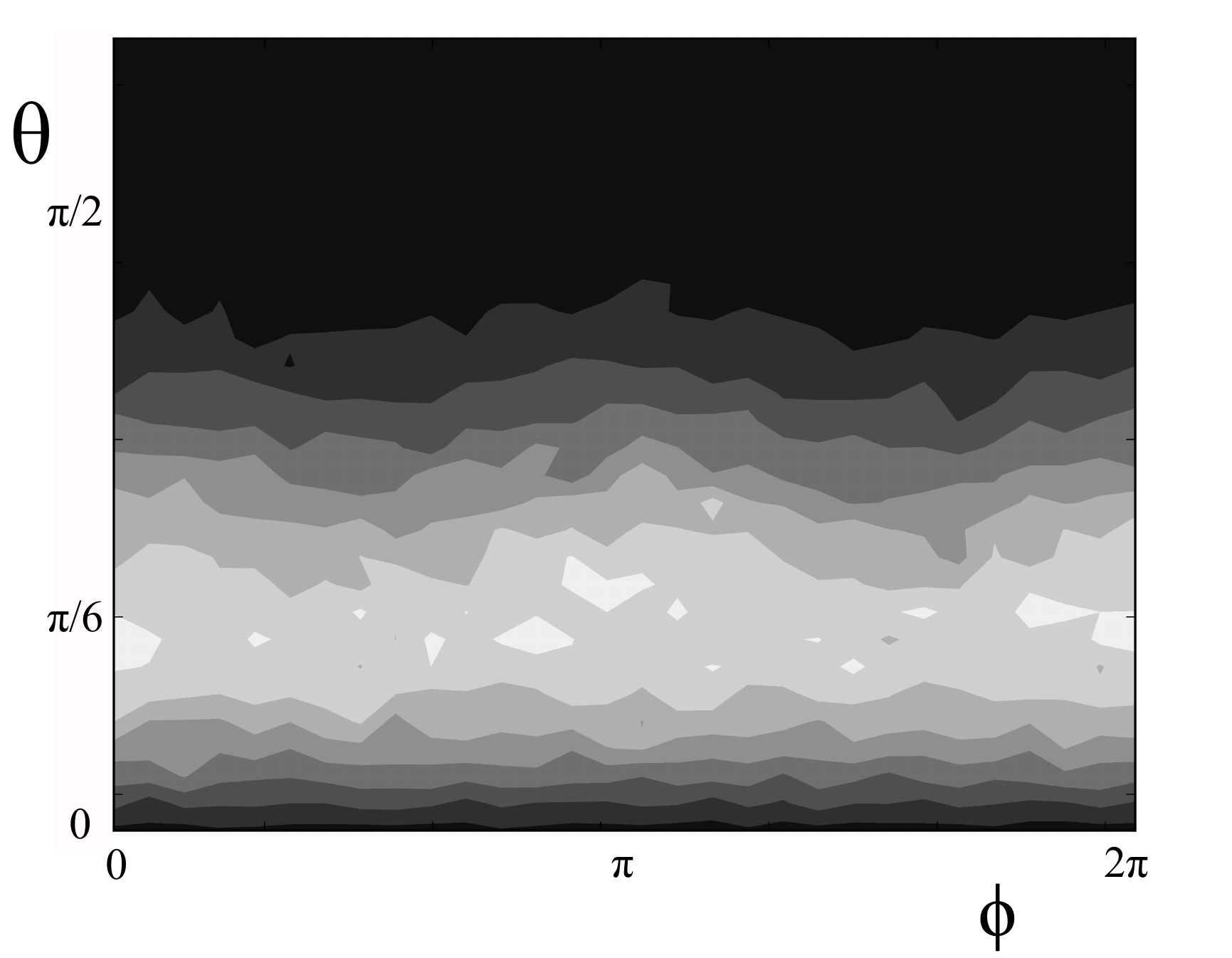}
\caption{$\theta-\phi$ profile of the directional signal generated from VLII simulation. Notice that the hint of hot spots at $\phi=0~[\pi]$ suggests the presence of anisotropies. \label{VLIIrate}}
\end{figure}

%The detector model and dark matter mass chosen here are the same as in Sec. ~\ref{Sec: VDF} but the lower energy threshold of the experiment has been lowered to $E_{\text{min}} = 2 \keV$. Indeed, the escape velocity of the VLII particles in the neighborhood of the Sun has been determined to be about $545$ km/s and the $v_{\text{low}}$ corresponding to $E_{\text{min}}=5\keV$ was then too high ($518$~km/s). With the new lower energy threshold, $v_{\text{low}}\approx 330$ km/s, which, although allowing more phase space, stays bigger than the Earth velocity. The nuclear recoils in the detector will then be almost collinear to the direction of the incoming dark matter particles. 

The $\theta-\phi$ profile of the directional detection rate generated is shown in Fig.~\ref{VLIIrate}. The hint of hot spots at $\phi=0~[\pi]$ suggest a suppression of circular orbits perpendicular to the plane of the Earth's orbit in the halo. 
%suggests an enhancement of circular orbits perpendicular to the plane of the Earth's orbit in the halo {\bf ?????????}
 At this point this is just a qualitative statement, which will be made quantitative below by the results of the fits.   Note that the plane of the Earth's orbit was chosen arbitrarily given that VLII does not have a baryonic disk.

%Due to the high value of $v_{\text{low}}$, the presence of tiny hotspots in $\phi=0\text{ }[\pi]$ indicates that the $x$-component of the dark matter particles velocity dominates over the $z$-component as for the Michie distribution introduced in Sec.~\ref{Sec: VDF}.

%The presence of hot spots in $\phi=0\text{ }[\pi]$ is a feature shared by a very wide class of distribution and does not provide much information about the details of the distribution function of the particles scattering in the detector. However, as shown in the previous parts, the shape of the dark matter DF can be determined by expanding it in series and fitting the associated differential detection rate to find the coefficients of the expansion.

%\subsubsection{The $E$, $L_t$, $L_z$ distributions in the neighborhood of the Sun}

%We use the velocities of the VLII dark matter particles as an input distribution function to perform a mock experiement. The mock signal is then fit to a DF in the form of Eq.~\ref{expansion}. The dark matter detector being constrained to stay on Earth, only the dark matter particles in the neighborhood of the Sun, so about $r_0=8$ kpc away from the center of the galaxy, have been selected to scatter. The VLII dark matter halo being approximately spherically symmetric in this region, the particles allowed to scatter in the detector have been randomly selected in a spherical shell of central radius $r_0$ and thickness $200$ pc.  

\begin{figure}
\includegraphics[scale=0.5]{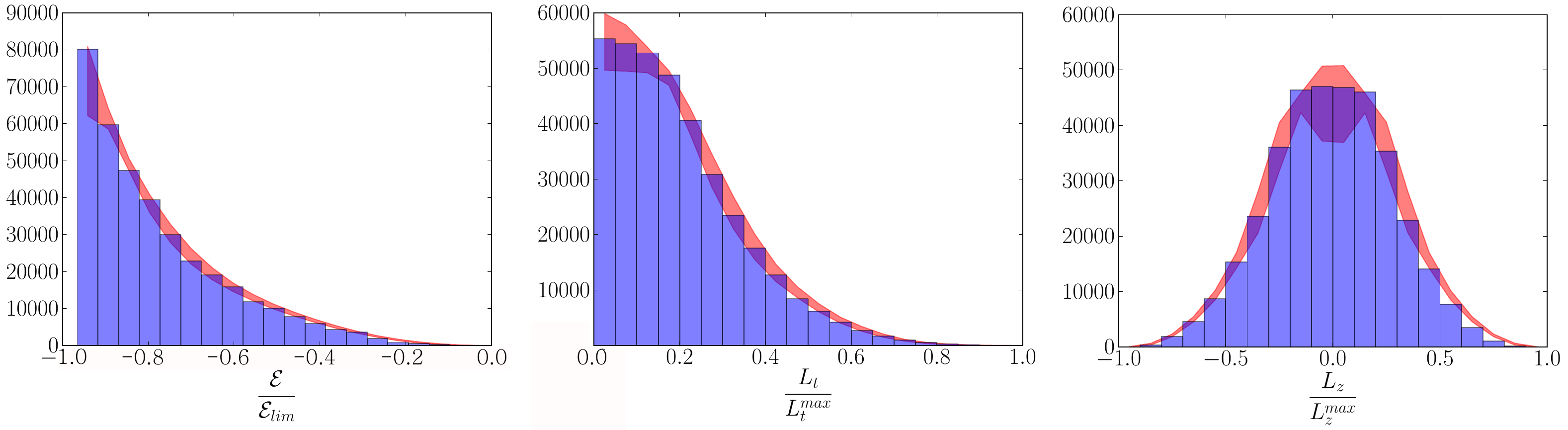}
\caption{Results of fits to a directional signal generated with VLII, displayed as distributions of $\mathcal{E}$, $L_y$ and $L_z$ for the particles kinematically allowed to scatter. The red band corresponds to the fitted distributions and its $1~\sigma$ error, whereas the blue histograms correspond the true underlying distributions.\label{Distr}}
\end{figure}

\begin{figure}
    \centering
\includegraphics[width=6.5in]{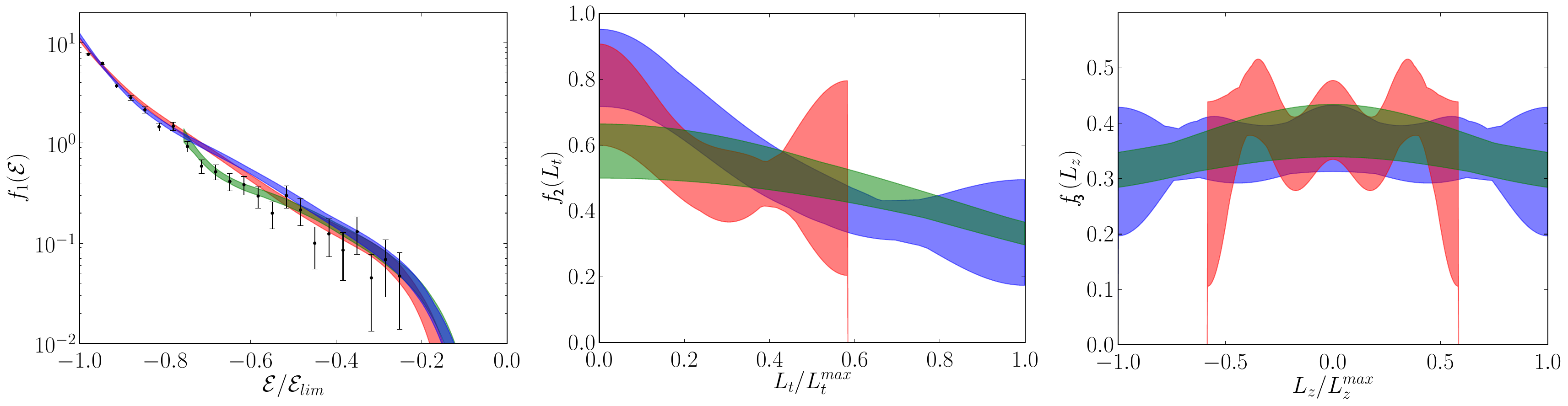}
\caption{Fits for (a) $f_1(\mathcal{E})$, (b) $f_2(L_t)$, and (c) $f_3(L_z)$ at 4.5 kpc (red), 8 kpc (blue) and 30 kpc (green) using the distribution of Via Lactea II. The black dots with error bars in (a) correspond to values of $f_1(\mathcal{E})$ extracted directly from data at $8$ kpc using Eq.~\ref{fEi}. The good agreement at different radii corroborates the use of Jeans Theorem. \label{Jeans}}
\end{figure}

We perform unbinned log-likelihood fits to this directional signal as described in Secs.~\ref{decomposition},\ref{Sec: VDF}. The results of the fits are displayed in Fig.~\ref{Distr}, where we plot $\mathcal{E}$, $L_t$ and $L_z$ distributions for the particles that are kinematically allowed to generate nuclear recoils within the recoil energy acceptance window. We contrast the distributions extracted from the fit with the true underlying distributions to show that the fit is correctly capturing the local VLII phase space distribution within errors. The fits for the $\mathcal{E}$, $L_t$ and $L_z$ dependent parts of the DF, denoted by $f_1(\mathcal{E})$, $f_2(L_t)$ and $f_3(L_z)$ in eq.~\ref{separable}, are displayed as blue bands in Fig.\ref{Jeans}. One can see that although $f_3(L_z)$ is consistent with a flat function, $f_2(L_t)$ is suppressed for high values of $L_t$, meaning that orbits perpendicular to the plane of the Earth's orbit in the halo are disfavored.  That confirms the presence of anisotropy mentioned earlier, and is consistent with the hot spots developed in the rate at $\phi=0~[\pi]$, see Fig.~\ref{VLIIrate}.

%According to Fig.~\ref{Distr}, parameterization in Eq.~\ref{expansion} for the dark matter DF seems to reproduce pretty well the behavior of the VLII dark matter particles in the neighborhood of the Sun. The increase in the width of the error bands near $L_t=0$ and $L_z=0$ is due to the fact that low velocity particles are less likely to be observed.

%Fitting the differential detection rate for $10^4$ signal events allows to find the one sigma error bands on the $E$, $L_t$ and $L_z$ distributions, compared against data in Fig.~\ref{Distr}. Of course, only particles with $v>v_{\text{low}}$, that is to say, able to be seen in the detector, contribute to the different distributions.

\subsection{Global Extrapolation of the Local Fits}

As discussed in Sec.~\ref{Sec: Jeans}, Jeans Theorem tells us if the dark matter distribution function can be expressed locally in terms of integrals of motion, then it should be valid at all other positions of the halo, provided that the distribution is in a steady state of equilibrium. In this section we test the validity of our choice of $\mathcal{E}$, $L_t$ and $L_z$ as integrals of motion by checking whether the VLII distribution function at other positions in the simulation agrees with our fit at 8 kpc.

%If $E$, $L_t$ and $L_z$ are integrals of motion, the DF fit through direct detection should not depend on the position of the observer in the galaxy. 

The most straightforward variable to check is $\mathcal{E}$. Consider the distribution of dark matter particles at a given location $\vec{r}$, binned in $\mathcal{E}$, $L_t$ and $L_z$ with bin widths $\Delta\mathcal{E}$, $\Delta L_t$ and $\Delta L_z$ respectively. The number of particles in the bin centered around $\mathcal{E}$, $L_{t}$ and $L_{z}$ is given by:
\begin{equation}
N(\mathcal{E},L_{t},L_{z})= \Delta\mathcal{E} \Delta L_t  \Delta L_z~\text{Det}[J(\mathcal{E},L_{t},L_{z})]~f_1(\mathcal{E})f_2(L_{t})f_3(L_{z}),
\end{equation}
where $\text{Det}[J(\mathcal{E},L_y,L_z)]$ is the determinant of the Jacobian, given by:
\begin{equation}
\text{Det}[J(\mathcal{E},L_t,L_z)]=\frac{1}{\sqrt{2\mathcal{E}+v_\text{esc}^2-L_t^2/r^2-L_z^2/r^2}}.
\end{equation}

Since the bins with better statistics are those with lower values of $\mathcal{E}$, and hence lower values of $L_{t}$ and $L_{z}$, we can extract $f_1(\mathcal{E})$ from bins where $L_{t}$ and $L_{z}$ are centered around zero and $\mathcal{E}$ is centered around $\mathcal{E}_i$:
\begin{equation}
\label{fE}
f_1(\mathcal{E}_i)=\frac{1}{f_2(0)f_3(0)}\frac{1}{\Delta\mathcal{E} \Delta L_t  \Delta L_z}\frac{N(\mathcal{E}_i,0,0)}{\sqrt{2\mathcal{E}_i+v_\text{esc}^2}}.
\end{equation}
\begin{figure}[t]
\includegraphics[scale=0.45]{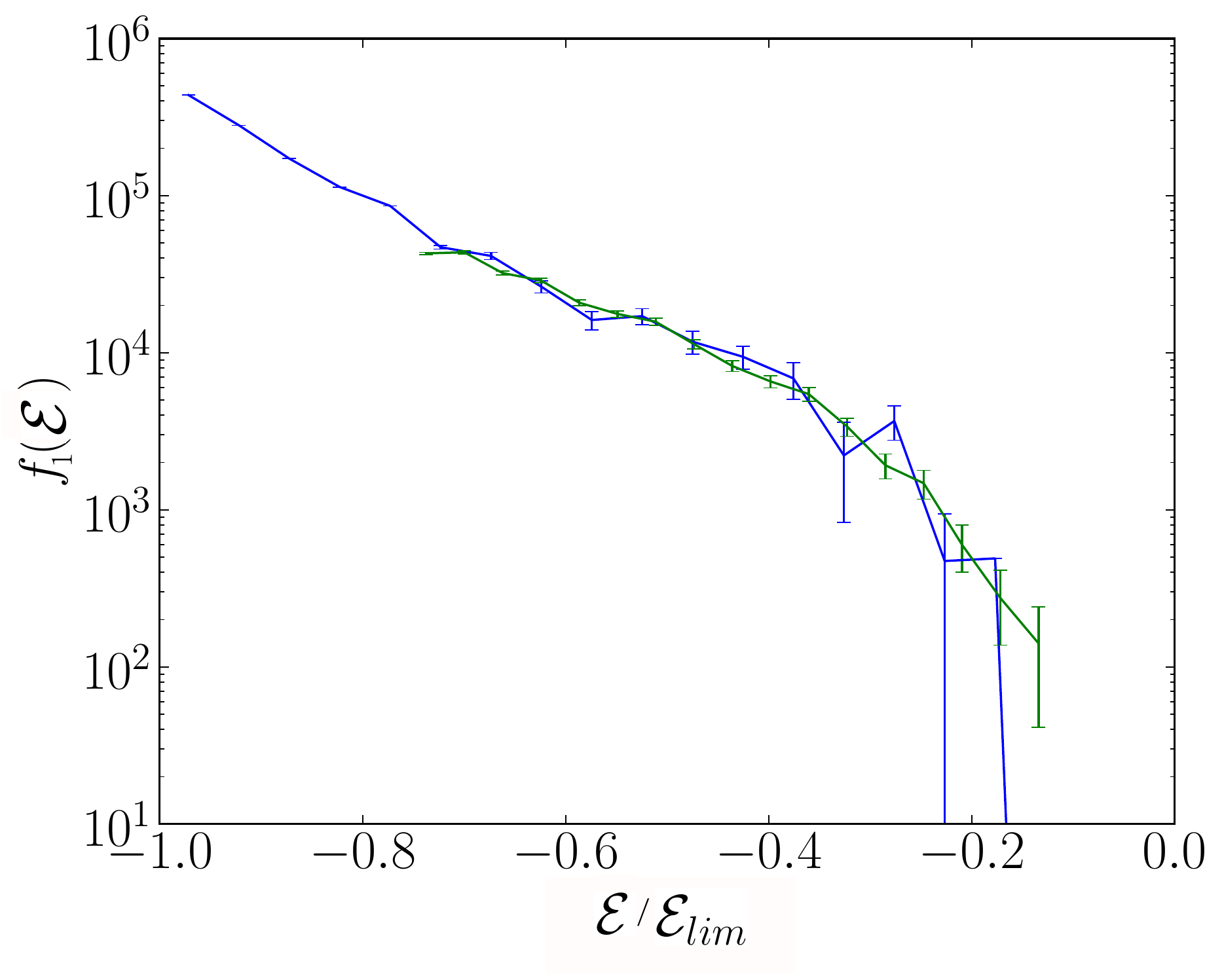}
\caption{$f_1(\mathcal{E})$ extracted from VLII data for $r\approx 8 \kpc$ (blue) and $r \approx 30 \kpc$ (green), using Eq.~\ref{fEi}. The values of $\mathcal{E}$ are normalized to $\mathcal{E}_\text{lim}(r)=-\psi(r)$ at $r=8\text{ kpc}$.  \label{jeansDATA}}
\end{figure}

The first factors in Eq.~\ref{fE} are independent of $\mathcal{E}$ and can be normalized away, finally giving:
\begin{equation}
\label{fEi}
f_1(\mathcal{E}_i)=\frac{N(\mathcal{E}_i,0,0)}{\sqrt{2\mathcal{E}_i+v_\text{esc}^2}}.
\end{equation}

In Fig.~\ref{jeansDATA}, we plot $f_1(\mathcal{E})$ using Eq.~\ref{fEi} for two different positions in the VLII simulation, namely $r=8\text{ kpc}$ and $r=30\text{ kpc}$. The distribution of particles at different radii are obtained from shells with the corresponding radii and 200 pc-thickness. Note that since the gravitational potential decreases with the distance to the center of the galaxy, the minimum energy of the particles at $r=30\text{ kpc}$ ($\mathcal{E}_{\text{min}}(r)=-\psi(r)$) is higher than at $r=8\text{ kpc}$. The excellent agreement between the extraction of $f_1(\mathcal{E})$ for the two radii reinforces the validity of Jeans Theorem and the fact the $\mathcal{E}$ is a good integral of motion. 
%\subsubsection{The velocity distribution function at other positions in the galaxy}

Unfortunately, there is not enough statistics to extract $f_2(L_t)$ and $f_3(L_z)$ directly from the simulation using the same method used above for $f_1(\mathcal{E})$, since the population in bins with increasing values of $L_t$ and $L_z$ rapidly drop in number. The alternative is to fit the dark matter distributions at different radii using our parameterization in Eqs.~\ref{separable} and \ref{expansion}. We perform such fits for three different radii, $r=4.5\text{ kpc}$, $r=8\text{ kpc}$ and $r=30\text{ kpc}$, where as before we select particles within shells of 200 pc-thickness and radius $r$.

%There is not enough statistics to extract $f_2(L_t)$ and $f_3(L_z)$ directly from the data because most dark matter particles have very low velocities. The shapes of these functions at different positions in the galaxy can be found however by directly fitting the dark matter velocity distributions using the expansion shown in Eq.~\ref{expansion}. For each $r$ at which the DF has to be computed, a fit is performed over $10^4$ particles randomly selected in a spherical shell of central radius $r$ and thickness $200 \text{ pc}$. Here, in order to study the robustness of the result found at $r_0$ closer to and further away from the center of the galaxy, $f_1$, $f_2$ and $f_3$ are computed at $r_1=30$ kpc and $r_2=4.5$ kpc. Recall that $r<r_1$ corresponds to the region where the integrals of motion associated to the dark matter VLII halo have to be close to $E$, $L_t$ and $L_z$ because of the approximate spherical symmetry of the gravitational potential.

%The results for $r=4.5 \kpc$ are pretty similar to the ones found in the neighborhood of the Sun. At $r=30 \kpc$, however, small discrepancies begin to appear for $f_1(E)$ and $f_2(L_z)$.

Fig.~\ref{jeansDATA} shows the results of the fits for $f_1(\mathcal{E})$, $f_2(L_t)$ and $f_3(L_z)$ and their corresponding error bands. The fact that there is a reasonable agreement over a wide range range of galactic radii is evidence that the parameterization motivated by Jeans theorem is valid and meaningful. From Fig.~\ref{jeansDATA} it is also interesting to note that, although $f_3(L_z)$ is consistent with a flat distribution, $f_2(L_t)$ indicates that the VLII distribution is anisotropic, and the angular momentum component along the Earth's motion is suppressed. Another interesting fact is that $f_1(\mathcal{E})$ is not consistent with a single exponential, as is the case for Maxwellian or Michie distributions. Indeed, the high velocity tail of $f_1(\mathcal{E})$ is harder than Maxwellian, as pointed out in \cite{Lisanti:2010qx}.

%however shows that the shapes of $f_1(E)$ at $8$ and $30$ kpc should be similar. The disagreement observed for $f_1(E)$ in Fig.~\ref{Jeans} is actually due to the fact that for $r\approx 8 \kpc$, events with $E\approx 0$ are very rare and then, the corresponding Sec. of the $f_1(E)$ curve is poorly fitted. 

%Concerning $f(L_z)$ however, no similar explanation can be found and the observed discrepancy is probably due to the fact that the dark matter halo starts being less and less spherical beyond $r_1$ and that the integrals of motions can no longer be approximated as being $E$, $L_t$ and $L_z$.

%The $L_z$ distribution fits less well. This could possibly be due to the fact that the dark matter halo starts being less and less spherical beyond $r_1$ and that the integrals of motions can no longer be approximated as being $E$, $L_t$ and $L_z$.

\section{Discussion}
\label{Sec: Discussion}
In this paper we discussed important aspects of directional detection that will be relevant once dark matter is detected in underground experiments. Directional experiments will probe the dynamics and kinematics in the dark matter interactions with nuclei and allow us for directly measure the dark matter phase space distribution in the solar neighborhood.

It will be important to reduce our bias in extracting information from directional data. With that in mind, we proposed to parametrize the local distribution function in terms of integrals of motion. If the dominant component of underlying dark matter distribution in the solar neighborhood is in equilibrium, expressing the DF in terms of integrals of motion will allow us to infer the DF at other positions in the halo. That can have a potentially large impact in studies of indirect signals, as well as implications for our understanding of the formation of the Milky Way halo.  There is a chance that the local dark matter distribution could have a sizable component that is in a non-equilibrium configuration such as streams of dark matter arising from tidally disrupted dark matter sub-halos.  These features can be readily identified in directional detection experiments and can be subtracted off \cite{Lisanti:2011as,Kuhlen:2012fz,Vergados:2012xn}.  

We also proposed a decomposition of the distribution function in special functions of integrals of motion. That allows for a more model independent extraction of the true phase space distribution. We illustrated our method by generating Monte Carlo signals in a hypothetical directional detector and fitting the coefficients of our decomposition. This study was performed with an analytical model of the halo, as well as using Via-Lactea II simulation. In the later, we also tested the results of our fit by extrapolating the fitted local distribution to other positions in the halo and comparing to the true distribution. We found that for radii less than $\sim 30$ kpc our extrapolation worked to a very good approximation.

The number of signal events necessary to detect non-standard features of the dark matter distribution obviously depend on the degree of such effects. For Via Lactea II we found that  \OO(1000) events were necessary to detect presence anisotropies and departures from a Maxwellian distribution.

%This article has proposed a parameterization of the dark matter phase space distribution function using integrals of motion. We have shown that decomposing the DF using a series expansion allows directional detection experiments to reconstruct a wide range of dark matter halo models. This parameterization and expansion of the dark matter phase space distribution

The parameterization and decomposition of the dark matter phase space distribution we have proposed may also have applications beyond directional detection. For instance, it could also allow a better understanding of the dark matter haloes obtained in N-body simulations. In particular, reconstructing the dark matter phase space distribution would allow us to compare the different existing N-body simulations to each other, studying for example how the anisotropies of the dark matter distribution function depend on the different parameters of the simulation, especially the presence of a baryonic disk.

In conclusion, this article has shown the power of directional detection experiments in the post-dark matter-discovery era. By using the results from directional detection experiments and N-body simulations, a better understanding of dark matter in the Milky Way can be obtained. 

\section*{Acknowledgements}

We thank Mariangela Lisanti, Robert Feldman, Louie Strigari and Tom Theuns for useful discussions. Special thanks to Michael Kuhlen for having given us access to the Via Lactea II N-body simulation full dataset and for his enlightening explanations of it. SEH and JGW are supported by the US DOE under contract number DE-AC02-76-SFO0515. SEH is supported by a Stanford Graduate Fellowship. JGW is supported by the DOE's Outstanding Junior Investigator Award. DSMA acknowledges the hospitality of the Aspen Center for Theoretical Physics where this work was partially completed. DSMA is supported by the US Department of Energy. Fermilab is operated by Fermi Research Alliance, LLC under Contract No. DE-AC02-07CH11359 with the US Department. 
\appendix

\section{Analytic Scattering Rates}
\label{appendix}

In this appendix we derive useful analytical expressions for the directional rate of nuclear recoils.
As discussed in Sec.~\ref{Sec: Rate}, the differential nuclear recoil rate is given by 
\begin{eqnarray}
\label{Eq: Rate2}
\frac{dR}{dE_Rd\Omega}\propto \int~ d\vec{v} ~f(\vec{v})~\delta\left[\vec{v}\cdot\hat{v}_R-\rho(E_R,\theta)\right]\Theta[v_\text{esc}^2 - v^2],
\end{eqnarray}
where
\begin{eqnarray}
    \rho(E_R,\theta) = v_\text{min}(E_R) - \vec v_\oplus\cdot\hat v_R = v_\text{min}(E_R) - v_\oplus\cos\theta.
\end{eqnarray}
Using
\begin{eqnarray*}
    \hat v_R &=& (x_R,y_R,z_R)\equiv(\sin\theta\cos\phi,\cos\theta,\sin\theta\sin\phi),
\end{eqnarray*}
and integrating over $v_x$ gives
\begin{eqnarray}
\label{Eq: IntRate}
\frac{dR}{dE_Rd\Omega}&\propto&\frac{1}{|x_R|} \int~ dv_y\;dv_z ~f(v_x[v_y,v_z],v_y,v_z)\Theta[v_\text{esc}^2-v_x[v_y,v_z]^2-v_y^2-v_z^2],
\end{eqnarray}
where
\begin{eqnarray}
v_x(v_y,v_z)&=&\frac{1}{x_R}(\rho(E_R,\theta)-y_Rv_y-z_Rv_z).
\end{eqnarray}
The domain of integration for the variables $v_y$ and $v_z$ is then set by the condition
\begin{eqnarray}
    v_z^2(z_R^2+x_R^2)+2v_zz_R(y_Rv_y-\rho(E_R,\theta))+(\rho(E_R,\theta)-y_Rv_y)^2-x_R^2(v_\text{esc}^2-v_y^2) <0.
\end{eqnarray}
The limits of integration on $v_z$ are then
\begin{eqnarray}
    v_{z\pm}(v_y) = \frac{1}{x_R^2+z_R^2}\left[z_R(\rho(E_R,\theta)-y_Rv_y)\pm\sqrt{\Delta}\right],
\end{eqnarray}
where
\begin{eqnarray}
    \Delta = x_R^2\left[(x_R^2+z_R^2)(v_\text{esc}^2-v_y^2)-(\rho(E_R,\theta)-v_yy_R)^2\right].
\end{eqnarray}
The condition $\Delta>0$ sets the limits for $v_y$:
\begin{eqnarray}
    v_{y\pm}=y_R\rho(E_R,\theta)\pm\sqrt{(1-y_R^2)(v_\text{esc}^2-\rho^2(E_R,\theta))}.
\end{eqnarray}
With the limits of integration defined, the only remaining constraint is 
\begin{eqnarray}
    v_\text{esc}^2-\rho^2(E_R,\theta)) > 0,
\end{eqnarray}
which requires that for a given recoil energy $E_R$ and recoil angle $\theta$, the minimum velocity of the incoming dark matter particles \emph{in the galactic frame} must be smaller than the dark matter escape velocity.

We then do the variable change 
\begin{eqnarray}
    V_y & = & v_y - y_R\rho(E_R,\theta) = v_y - v_{y}^0,\\
    V_z & = & \frac{x_R^2+z_R^2}{|x_R|}v_z - \frac{z_R}{|x_R|}[\rho(E_R,\theta)(1-y_R^2)-y_RV_y] = \frac{x_R^2+z_R^2}{|x_R|}v_z - v_{z}^0.
\end{eqnarray}
The new boundaries are then given by $\pm V_{z}^0(V_y)$ and $\pm V_{y}^0$ with
\begin{eqnarray}
    V_{z}^0(V_y) &=& \sqrt{\Delta_0^2-V_y^2},\\
    V_{y}^0 &=& \Delta_0,
\end{eqnarray}
where
\begin{eqnarray}
    \Delta_0^2 = (1-y_R^2)(v_\text{esc}^2-\rho^2(E_R,\theta)).
\end{eqnarray}
The differential rate then becomes
\begin{eqnarray}
\label{mess1}
\nonumber\frac{dR}{dE_Rd\Omega}&\propto&\frac{1}{x_R^2+z_R^2}\int_{-\Delta_0}^{\Delta_0}\int_{-\sqrt{\Delta_0^2-V_y^2}}^{\sqrt{\Delta_0^2-V_y^2}}\Theta[v_\text{esc}^2-\rho^2(E_R,\theta)]\;dV_z\;dV_y\\
&\times& f\left[v_x\left(V_y+v_y^0,\frac{|x_R|}{x_R^2+z_R^2}(V_z + v_z^0)\right),V_y+v_y^0,\frac{|x_R|}{x_R^2+z_R^2}(V_z+v_z^0)\right].
\end{eqnarray}

For a distribution function expressed in terms of energy $\mathcal{E}$ and the components of the angular momentum $L_t$ and $L_z$,
%\begin{eqnarray}
%    f(\mathcal{E},L_t,L_z)=\sum_{i,j,k}c_{ijk} P_l^{(i)}(\mathcal{E})\cos(j\pi L_t)\cos(k\pi L_z)
%\end{eqnarray}
%it is possible to compute the differential detection rate analytically for each of the expansion terms.
we can recover the dependence on $V_y$ and $V_z$ using 
\begin{eqnarray}
    \mathcal{E}&=&\frac{1}{2}\left(V_y^2+V_z^2-\Delta_0^2\right),\\
    L_t &=& r_\oplus \frac{|x_r|}{x_r^2+z_r^2}(V_z+v_z^0),\\
    \label{mess2}
    L_z &=& r_\oplus (V_y+v_y^0).
    \end{eqnarray}
    
Eqs.~\ref{mess1} to \ref{mess2} allow us to integrate the rate analytically for several special forms of $f(\mathcal{E},L_t,L_z)$, such as
\begin{eqnarray}
\label{f1}\mathcal{E}^\ell~\cos(n\pi L_z)~\cos(m\pi L_t),\\
\mathcal{E}^\ell~\cos(n\pi L_z)~\sin(m\pi L_t),\\
\mathcal{E}^\ell~\sin(n\pi L_z)~\cos(m\pi L_t),\\
\mathcal{E}^\ell~\sin(n\pi L_z)~\sin(m\pi L_t).
\end{eqnarray}
For instance, for $f(\mathcal{E},L_t,L_z)$ as in Eq.~\ref{f1}, the differential rate is given by
\begin{eqnarray}
    \begin{split}
        \frac{dR}{dE_Rd\Omega}\propto&\frac{1}{2^l(x_R^2+z_R^2)}\int_{-\Delta_0}^{\Delta_0}\int_{-\sqrt{\Delta_0^2-V_y^2}}^{\sqrt{\Delta_0^2-V_y^2}} [V_y^2+V_z^2-\Delta_0^2]^\ell\\
    &\cos\left[m\pi r_\oplus\frac{|x_r|}{x_r^2+z_r^2}(V_z+v_z^0)\right]\cos\left[n\pi r_\oplus(V_y+v_y^0)\right]dV_z\;dV_y.
\end{split}
\end{eqnarray}

The integration above is straightforward to perform and we finally obtain:
\begin{align*}
\int\delta\left(\vec{v}\cdot\vec{v}_R-\rho\right)d^3v~\mathcal{E}^\ell\cos(n\pi L_z)\cos(m\pi L_t)=&\\
(-1)^\ell~ \ell!~ \pi~&\cos[\pi\rho(n\hat{y}_R+m\hat{z}_R)]\times\text{M}_+\\
+(-1)^\ell~ \ell!~ \pi~&\cos[\pi\rho(n\hat{y}_R-m\hat{z}_R)]\times\text{M}_-,
\end{align*}
where we set $r_\oplus=1$ to lighten the notation and
\begin{align*}
&\rho=v_E \cos\theta+v_\text{min}(E_R),\\
&\hat{v}_R\equiv(\hat{x}_R,\hat{y}_R,\hat{z}_R)\equiv(\sin\theta\cos\phi,\cos\theta,\sin\theta\sin\phi),
\\
\\
\text{M}_+=&\left(\frac{\sqrt{v_\text{esc}^2-\rho^2}}{\pi\sqrt{n^2+m^2-(n\hat{y}_R+m\hat{z}_R)^2}}\right)^{\ell+1}
J_{\ell+1}(\pi\sqrt{n^2+m^2-(n\hat{y}_R+m\hat{z}_R)^2}\sqrt{v_\text{esc}^2-\rho^2}),\\
\text{M}_-=&\left(\frac{\sqrt{v_\text{esc}^2-\rho^2}}{\pi\sqrt{n^2+m^2-(n\hat{y}_R-m\hat{z}_R)^2}}\right)^{\ell+1}
J_{\ell+1}(\pi\sqrt{n^2+m^2-(n\hat{y}_R-m\hat{z}_R)^2}\sqrt{v_\text{esc}^2-\rho^2}).
\end{align*}

Similar expressions hold for the others:
\begin{align*}
\int\delta\left(\vec{v}\cdot\vec{v}_R-\rho\right)d^3v~\mathcal{E}^\ell\sin(n\pi L_z)\cos(m\pi L_t)=&\\
(-1)^\ell~ \ell!~ \pi~&\sin[\pi\rho(n\hat{y}_R+m\hat{z}_R)]\times\text{M}_+\\
+(-1)^\ell~ \ell!~ \pi~&\sin[\pi\rho(n\hat{y}_R-m\hat{z}_R)]\times\text{M}_-,\\
\end{align*}

\begin{align*}
\int\delta\left(\vec{v}\cdot\vec{v}_R-\rho\right)d^3v~\mathcal{E}^\ell\cos(n\pi L_z)\sin(m\pi L_t)=&\\
(-1)^\ell~ \ell!~ \pi~&\sin[\pi\rho(n\hat{y}_R+m\hat{z}_R)]\times\text{M}_+\\
+(-1)^\ell~ \ell!~ \pi~&\sin[\pi\rho(-n\hat{y}_R+m\hat{z}_R)]\times\text{M}_-,\\
\end{align*}

\begin{align*}
\int\delta\left(\vec{v}\cdot\vec{v}_R-\rho\right)d^3v~\mathcal{E}^\ell\sin(n\pi L_z)\sin(m\pi L_t)=&\\
(-1)^\ell~ \ell!~ \pi~&(-\cos[\pi\rho(n\hat{y}_R+m\hat{z}_R)])\times\text{M}_+\\
+(-1)^\ell~ \ell!~ \pi~&\cos[\pi\rho(n\hat{y}_R-m\hat{z}_R)]\times\text{M}_-.\\
\end{align*}

Note that all those expressions can be written in a compact form:

\begin{align*}
\int\delta\left(\vec{v}\cdot\vec{v}_R-\rho\right)d^3v~\mathcal{E}^\ell~e^{i\pi(n L_z+m L_t)}~~=~~(-1)^\ell~ \ell!~ 2\pi~&e^{i\pi\rho(n\hat{y}_R+m\hat{z}_R)}\\
\times\left(\frac{\sqrt{v_\text{esc}^2-\rho^2}}{\pi\sqrt{n^2+m^2-(n\hat{y}_R+m\hat{z}_R)^2}}\right)&^{\ell+1}~
J_{\ell+1}(\pi\sqrt{n^2+m^2-(n\hat{y}_R+m\hat{z}_R)^2}\sqrt{v_\text{esc}^2-\rho^2}).\\
\end{align*}

Remembering that
\begin{align*}
\hat{v}_R&\equiv(\hat{x}_R,\hat{y}_R,\hat{z}_R),\\
\vec{L}&\equiv(0,L_z,L_t),
\end{align*}

we can rewrite the expression above in a frame independent form:

\begin{align*}
\int\delta\left(\vec{v}\cdot\vec{v}_R-\rho\right)d^3v~\mathcal{E}^\ell~e^{i\vec{L}\cdot\vec{u}_T}~~=~~(-1)^\ell~ \ell!~ 2\pi~e^{i\hat{v}_R\cdot\vec{u}_T}
~(v_\text{esc}^2-\rho^2)^{\ell+1}~\frac{J_{\ell+1}(|\vec{u}_T|\Xi)}{(|\vec{u}_T|\Xi)^{\ell+1}},
\end{align*}
\begin{align*}
\text{with}\quad\Xi\equiv\sqrt{v_\text{esc}^2-\rho^2}\sqrt{1-(\hat{v}_R\cdot\hat{u}_T)^2},
\end{align*}

where $\vec{u}_T$ is an arbitrary vector in the tangential plane.

Finally, when there is no dependence on $\vec{L}$ the rate takes the form:

\begin{align*}
\int\delta\left(\vec{v}\cdot\vec{v}_R-\rho\right)d^3v~\mathcal{E}^\ell~~=~~2\pi~\frac{(-1)^\ell}{\ell+1}\left(\frac{v_\text{esc}^2-\rho^2}{2}\right)^{\ell+1}.\\
\end{align*}

In particular, for a generic $f(\mathcal{E})$,

\begin{equation*}
\int\delta\left(\vec{v}\cdot\vec{v}_R-\rho\right)d^3v~f(\mathcal{E})~~=~~2\pi\int_{-\left(\frac{v_\text{esc}^2-\rho^2}{2}\right)}^{0}f(\mathcal{E})~d\mathcal{E}.
\end{equation*}

\bibliography{DarkMatter}

\end{document}